\documentclass[12pt]{article}
\usepackage{amsfonts}
\usepackage{amsmath}
\usepackage{amssymb}
\usepackage{amsthm}
\usepackage{accents}
\usepackage{youngtab}
\usepackage{braket}
\usepackage{graphicx}
\usepackage[svgnames]{xcolor}
\usepackage{here}
\usepackage{comment}
\usepackage{simplewick}
\usepackage{tikz}
\usetikzlibrary{decorations.markings}
\usepackage{feynmf}

\newcommand{\ba}{\begin{eqnarray}}
\newcommand{\ea}{\end{eqnarray}}
\newcommand{\nn}{\nonumber}

\newcommand{\cA}{\mathcal{A}}
\newcommand{\cB}{\mathcal{B}}
\newcommand{\cC}{\mathcal{C}}
\newcommand{\cD}{\mathcal{D}}
\newcommand{\cN}{\mathcal{N}}

\newcommand{\lt}{\left}
\newcommand{\rt}{\right}

\tikzset{->-/.style={decoration={
  markings,
  mark=at position .55 with {\arrow{stealth}}},postaction={decorate}}}

\tikzset{-<-/.style={decoration={
  markings,
  mark=at position .55 with {\arrow[>=stealth]{<}}},postaction={decorate}}}  

 \DeclareMathOperator*{\Res}{Res}

\makeatletter
\newcommand{\douwidehat}[2]{%
  \sbox0{$\m@th#1\widehat{\hphantom{#2}}$}%
  \sbox2{$\m@th#1x$}
  \sbox4{$\m@th#1#2$}
  \dimen0=\ht0
  \advance\dimen0 -.8\ht2
  \dimen2=\dp4
  \rlap{%
    \raisebox{\dimexpr\dimen0-\dimen2}{%
      \scalebox{1}[-1]{\box0}%
    }%
  }%
  {#2}%
}
\makeatother

 \makeatletter
\@addtoreset{equation}{section}

\makeatother

\usepackage[colorlinks=true,linkcolor=black,citecolor=black,urlcolor=black]{hyperref}

\begin{document}

\begin{titlepage}

\begin{center}
{\Large
{\bf  Bethe/Gauge Correspondence\\[.2em] for\\[.5em] SO/Sp Gauge Theories and Open Spin Chains}
}
\end{center}

\vskip 1cm

\begin{center}
{\Large
Taro Kimura$^\dagger$ and Rui-Dong Zhu$^{\ddagger,\ast}$
}
\end{center}

\vskip 1 cm

\begin{center}
$^\dagger$ {\it Institut de Math\'ematiques de Bourgogne,\\ Universit\'e Bourgogne Franche-Comt\'e, 21078 Dijon, France}\\
$^\ddagger$ {\it Institute for Advanced Study \& School of Physical Science and Technology,\\ Soochow University, Suzhou 215006, China}\\
$^\ast$ {\it School of Theoretical Physics, Dublin Institute for Advanced Studies\\ 10 Burlington Road, Dublin, Ireland}
\end{center}

\vskip 3cm

\begin{abstract}
In this article, we extend the work of \cite{NS-2} to a Bethe/Gauge correspondence between 2d (or resp. 3d) SO/Sp gauge theories and open XXX (resp. XXZ) spin chains with diagonal boundary conditions. The case of linear quiver gauge theories is also considered. 
\end{abstract}

\end{titlepage}

\noindent\hrulefill
\tableofcontents
\noindent\hrulefill

\section{Introduction}

The integrable nature of supersymmetric gauge theories with eight super charges has gathered intensive interests since the groundbreaking work of Seiberg and Witten \cite{SW,Seiberg:1994aj}. In 4d, an R-matrix associated to the instanton counting was discovered on the full $\Omega$-background \cite{MO,Nakajima2013}, and such an R-matrix is known to be attached to an algebra constructed by coproducting the affine Yangians of symmetric Kac-Moody Lie algebras. When the gauge group of the gauge theory is $G=A_n$, the underlying algebra is a coproduct of $n$ copies of the affine Yangian of $\mathfrak{gl}_1$ \cite{MO,SHc}. The affine Yangian of $\mathfrak{gl}_1$ can be further uplifted to a K-theoretic version known as the Ding-Iohara-Miki algebra (or the quantum toroidal algebra of ${\mathfrak{gl}}_1$) \cite{DI,Miki,FFJMM1} and its elliptic version, the elliptic Ding-Iohara-Miki algebra \cite{Saito}. Through these uplifted algebras, we can figure out the action of the algebra \cite{AFS,Zhu-elliptic} and the R-matrix \cite{FJMM1,Awata:2017lqa} on the 5d $\cN=1$ gauge theories and a special class of 6d $\cN=(1,0)$ theories, which can be beautifully presented in the brane web picture of the gauge theories \cite{AHK,M-strings}. However, the quantum toroidal algebra acting on the gauge theories with gauge groups other than the A-type ones is yet to be clarified (see e.g. footnote 8 in \cite{KZ-Zn}). It is thus still not clear whether the integrability holds on the full $\Omega$-background (for example we refer to the complicated expressions of the qq-characters of BCD-type gauge groups \cite{Haouzi:2020yxy} in contrast to A-type gauge groups \cite{BPS/CFT,Kimura-Pestun,Bourgine:2015szm,Kimura:2016dys,5dBMZ}).
The correspondence has been studied for arbitrary $G$ between $G$-SYM and $\widehat{{}^L{G}}$-Toda chain in the absence of the $\Omega$-background~\cite{Martinec:1995by}, and there are some attempts \cite{Kimura:2017hez,Kimura:2019gon} to build the corresponding algebra from the Nekrasov partition function of the BCD-type quiver gauge theories.\footnote{There is a fiber-base duality connecting the gauge theory with gauge group $G_1$ and quiver structure $\Gamma_1$ to the theory with gauge group $G_2=\Gamma_1$ and quiver structure $\Gamma_2=G_1$ \cite{fiber-base} for $G_{1,2}$ and $\Gamma_{1,2}$ being ADE type. The situation is expected to be more involved for the non-simply-laced cases. This duality was also discussed in the algebraic approach in \cite{Bourgine:2018fjy}.} 
Nevertheless, the structure of the quantum algebra is still far from clear at the current stage.

In this work, we take a bottom-up approach to study the integrability of the gauge theories with BCD-type gauge groups by generalizing the Bethe/Gauge correspondence proposed in \cite{NS-2,NS1,NS2}. The original statement was about the duality between 2d $\cN=(2,2)^\ast$ (or resp. 3d $\cN=2^\ast$; 4d $\cN=1^\ast$) SU($N$) gauge theories with XXX (resp. XXZ; XYZ) spin chains. A more modern understanding of the relation with 4d $\cN=2$ (resp. 5d $\cN=1$; 6d $\cN=(1,0)$) gauge theories is explained in~\cite{Dorey:2011pa,Chen:2011sj,Chen:2012we} in details by Higgsing the theory in the so-called Nekrasov-Shatashvili (NS) limit to obtain vortex strings in the Higgs phase. See also ealier related works~\cite{Dorey:1998yh,Dorey:1999zk}. In this description, the 2d $\cN=(2,2)^\ast$ (or resp. 3d $\cN=2^\ast$; 4d $\cN=1^\ast$) theory that captures the integrability nature is an effective theory on the worldvolume of the vortex strings. In relation to the underlying algebra that hosts the R-matrix,%
\footnote{A possible gauge theory interpretation of the R-matrix in the context of the Bethe/Gauge correspondence has been addressed in~\cite{Bullimore:2017lwu}.
}
we remark that a similar (rescaled) NS limit can take the quantum toroidal algebra of ${\mathfrak{gl}}_1$ to the quantum group $U_q(\widehat{\mathfrak{sl}}_2)$, whose finite dimensional representations are known to give the solutions to the $RTT$-relation of the six-vertex model (XXZ spin chain). In this article, we study the correspondence between 2d (or 3d) gauge theories with SO or Sp gauge groups and XXX (or XXZ) spin chain with open boundary conditions. We will also briefly discuss the relation between the results obtained here and the string-theory set-up used in the derivation of the Bethe/Gauge correspondence in the A-type case. 

This article is organized as follows. In section \ref{s:R-mat}, we give a brief review on the integrability of the closed and open XYZ spin chain. In section \ref{s:gauge-th}, we review on the $D^2$ ($\times S^1$) partition function of 2d $\cN=(2,2)$ (3d $\cN=2$) gauge theory, and write down the effective potential. By using this effective potential, we first reproduce the well-known Bethe/Gauge correspondence between the vacuum equation of the A-type gauge theories and the Bethe ansatz equation of the closed XXX (or XXZ) spin chains in section \ref{s:dic-A}, and extend this duality to the case of BCD-type gauge theories and open spin chains in section \ref{s:corr-open}. We further push forward the computation to the $A_2$ quiver gauge theories in section \ref{s:A2-qui}, to a general linear quiver in section \ref{s:Ar_quiver}, and discuss some potential physical meanings of our results in the context of the string theory in section \ref{s:dis}.

\section{The R-matrix and Integrable Spin Chains}\label{s:R-mat}

The integrability of a spin chain is characterized by an R-matrix, ${\bf R}(u):\ V\otimes V\rightarrow V\otimes V$, satisfying the Yang-Baxter equation, 
\ba
{\bf R}_{12}(u-v){\bf R}_{13}(u){\bf R}_{23}(v)={\bf R}_{23}(v){\bf R}_{13}(u){\bf R}_{12}(u-v).\label{YBE}
\ea
The most general R-matrix for a solvable spin-$\frac{1}{2}$ $\mathfrak{sl}_2$-XYZ spin chain model is given by \cite{baxter2007exactly}
\ba
{\bf R}(u)=\lt(\begin{array}{cccc}
\alpha(u) & & & \delta(u)\\
& \beta(u) & \gamma(u) & \\
& \gamma(u) & \beta(u) & \\
\delta(u) & & & \alpha(u)\\
\end{array}\rt),\label{R-XYZ}
\ea
where
\begin{subequations}
\ba
\alpha(u)=\frac{\theta_{0,1/2}(u,2\tau)\theta_{1/2,1/2}(u+\eta,2\tau)}{\theta_{0,1/2}(0,2\tau)\theta_{1/2,1/2}(\eta,2\tau)},\quad \beta(u)=\frac{\theta_{1/2,1/2}(u,2\tau)\theta_{0,1/2}(u+\eta,2\tau)}{\theta_{0,1/2}(0,2\tau)\theta_{1/2,1/2}(\eta,2\tau)},\\
\gamma(u)=\frac{\theta_{0,1/2}(u,2\tau)\theta_{0,1/2}(u+\eta,2\tau)}{\theta_{0,1/2}(0,2\tau)\theta_{0,1/2}(\eta,2\tau)},\quad \delta(u)=\frac{\theta_{1/2,1/2}(u,2\tau)\theta_{1/2,1/2}(u+\eta,2\tau)}{\theta_{0,1/2}(0,2\tau)\theta_{0,1/2}(\eta,2\tau)},
 \ea
\end{subequations}
with 
\ba
\theta_{a_1,a_2}(u,\tau)=\sum_{m=-\infty}^\infty \exp\lt(i\pi \lt((m+a_1)^2\tau+2(m+a_1)(u+a_2)\rt)\rt).
\label{theta_fn}
\ea

Let us list several useful properties of the above R-matrix. 
\begin{itemize}
\item ${\bf R}(0)={\cal P}$, where ${\cal P}$ is the permutation operator that acts as ${\cal P}(x\otimes y)=y\otimes x$ for $^\forall x,y\in V$. 

\item ${\bf R}_{21}(u)={\bf R}_{12}(u)={\bf R}_{12}(u)^{t_1t_2}$, where $t_i$ stands for the transpose in the $i$-th vector space. 

\item Unitarity: ${\bf R}_{12}(u){\bf R}_{21}(-u)={\bf R}_{12}(u){\bf R}_{12}(-u)=-\frac{\sigma(u-\eta)\sigma(u+\eta)}{\sigma(\eta)^2}{\bf I}=:\rho(u){\bf I}$, for 
\ba
\sigma(u)=\theta_{1/2,1/2}(u,\tau).\label{sigma-def}
\ea
Note that we used 
\ba
\theta_{0,\frac{1}{2}}^2(x)\theta_{0,\frac{1}{2}}^2(y)-\theta_{\frac{1}{2},\frac{1}{2}}^2(x)\theta_{\frac{1}{2},\frac{1}{2}}^2(y)=\theta_{0,\frac{1}{2}}(x+y)\theta_{0,\frac{1}{2}}(x-y)\theta^2_{0,\frac{1}{2}}(0),
\ea
and 
\ba
\theta^2_{\frac{1}{2},\frac{1}{2}}(x)\theta^2_{0,\frac{1}{2}}(y)-\theta^2_{0,\frac{1}{2}}(x)\theta^2_{\frac{1}{2},\frac{1}{2}}(y)=\theta_{\frac{1}{2},\frac{1}{2}}(x+y)\theta_{\frac{1}{2},\frac{1}{2}}(x-y)\theta_{0,\frac{1}{2}}^2(0),
\ea
in the derivation. 

\item Crossing unitarity: ${\bf R}_{12}(u)=V_1{\bf R}^{t_2}_{12}(-u-\eta)V_1$ for $V=-i\sigma_y$, so we have 
\ba
{\bf R}_{12}^{t_1}(u){\bf R}_{12}^{t_1}(-u-2\eta)=V_1{\bf R}_{12}(u+\eta)V_1^2{\bf R}_{12}(-u-\eta)V_1=\rho(u-\eta){\bf I}\nn\\
=-\frac{\sigma(u+2\eta)\sigma(u)}{\sigma(\eta)^2}{\bf I}=:\rho'(u){\bf I}.\label{ref-uni}
\ea

\end{itemize}

A concrete integrable model is then given by the monodromy matrix, ${\bf T}(u)\in {\rm End}(V^{(0)}\otimes V^{\otimes L})$, and the transfer matrix, $\mathsf{t}(u)\in {\rm End}(V^{\otimes L})$ built from the R-matrix.\footnote{$V^{(0)}$ is called the auxiliary quantum space. One can certainly take $L$ vector spaces, on which the transfer matrix act, to be different. A typical choice is to take different representation of the R-matrix at different site with spin $s_i$.}
The most well-studied model is the closed spin chain with periodic boundary condition, whose monodromy matrix is given by  
\ba
{\bf T}_0(u)={\bf R}_{0L}(u-\vartheta_L)\dots {\bf R}_{01}(u-\vartheta_1),\label{mono-mat}
\ea
where $\vartheta_i$'s are called the inhomogeneous parameters in the closed spin chain.\footnote{Not to be confused with the theta functions~\eqref{theta_fn}.} The monodromy matrix satisfies the RTT-relation, 
\ba
{\bf R}_{00'}(u-v){\bf T}_0(u){\bf T}_{0'}(v)={\bf T}_{0'}(v){\bf T}_0(u){\bf R}_{00'}(u-v),\label{RTT}
\ea
which directly follows from the Yang-Baxter equation (\ref{YBE}). The transfer matrix is then given by 
\ba
\mathsf{t}(u)={\rm tr}_0\lt({\bf T}_0(u)\rt),
\ea
and one can show by using (\ref{RTT}) that  
\ba
\lt[\mathsf{t}(u),\mathsf{t}(u')\rt]=0,\quad {\rm for}\ ^\forall u,u'.\label{com-pro-transfer}
\ea

Following the property (\ref{com-pro-transfer}), it is easy to see that all the charges defined as the expansion coefficients of the transfer matrix, 
\ba
\mathsf{t}(u)=:\sum_{n=0}^\infty H^{(n)}u^n,
\ea
commute with each other, i.e. $\lt[H^{(n)},H^{(m)}\rt]=0$ for $^\forall m,n$. These charges characterize an integrable system described by the Hamiltonian 
\ba
{\cal H}:=\frac{1}{H^{(0)}}H^{(1)}.\label{d-Ham}
\ea

The corresponding Hamiltonian for the closed spin-$\frac{1}{2}$ XYZ chain constructed from (\ref{R-XYZ}) (with all inhomogeneous parameters turned off) is given by 
\ba
{\cal H}=\frac{1}{2}\sum_{n=1}^L\lt(J_x\sigma^{(n)}_x\sigma^{(n+1)}_x+J_y\sigma^{(n)}_y\sigma^{(n+1)}_y+J_z\sigma^{(n)}_z\sigma^{(n+1)}_z\rt),
\ea
where $\sigma_{x,y,z}^{(n)}$ are the Pauli matrices assigned to the site $n$, and the couplings are parametrized as follows:
\ba
J_x=e^{i\pi \eta}\frac{\sigma(\eta+\frac{\tau}{2})}{\sigma(\frac{\tau}{2})},\quad J_y=e^{i\pi \eta}\frac{\sigma(\eta+\frac{1+\tau}{2})}{\sigma(\frac{1+\tau}{2})},\quad J_z=\frac{\sigma(\eta+\frac{1}{2})}{\sigma(\frac{1}{2})},
\ea
and $\sigma(u)$ is defined in (\ref{sigma-def}).

\subsection{Reduction to XXZ chain}

The XXZ limit can be taken by setting $\tau\rightarrow i\infty$, 
and the following rewriting of the $\theta$-functions is useful to take the limit,
\begin{subequations}
\ba
&&\theta_{0,0}(u,\tau)=\prod_{m=1}^\infty (1-q^{2m})(1+e^{2\pi iu}q^{2m-1})(1+e^{-2\pi iu}q^{2m-1}),\label{theta-prod-1}\\
&&\theta_{0,1/2}(u,\tau)=\prod_{m=1}^\infty (1-q^{2m})(1-e^{2\pi iu}q^{2m-1})(1-e^{-2\pi iu}q^{2m-1}),\label{theta-prod-2}\\
&&\theta_{1/2,0}(u,\tau)=2q^{\frac{1}{4}}\cos(\pi u)\prod_{m=1}^\infty (1-q^{2m})(1+e^{2\pi iu}q^{2m})(1+e^{-2\pi iu}q^{2m}),\label{theta-prod-3}\\
&&\theta_{1/2,1/2}(u,\tau)=-2q^{\frac{1}{4}}\sin(\pi u)\prod_{m=1}^\infty (1-q^{2m})(1-e^{2\pi iu}q^{2m})(1-e^{-2\pi iu}q^{2m}),\label{theta-prod-4}
\ea
\end{subequations} 
where we set $q:=e^{\pi i\tau}$. In the XXZ limit, $q\rightarrow 0$, and we see that 
\begin{subequations}
\ba
&&\theta_{0,0}(u,\tau)\rightarrow 1,\quad \theta_{0,1/2}(u,\tau)\rightarrow 1,\\
&&\theta_{1/2,0}(u,\tau)\sim 2q^{\frac{1}{4}}\cos(\pi u),\quad \theta_{1/2,1/2}(u,\tau)\sim -2q^{\frac{1}{4}}\sin(\pi u).
\ea
\end{subequations}
Therefore we have 
\begin{subequations}
\ba
J_x\sim e^{i\pi\eta}\frac{\sin(\pi(\eta+\frac{\tau}{2}))}{\sin(\frac{\pi\tau}{2})}\sim e^{i\pi\eta}\frac{\exp(-i\pi(\eta+\frac{\tau}{2}))}{\exp(-i\frac{\pi\tau}{2})}\rightarrow 1,\\
J_y\sim e^{i\pi\eta}\frac{\sin(\pi(\eta+\frac{1+\tau}{2}))}{\sin(\frac{\pi(1+\tau)}{2})}\sim e^{i\pi\eta}\frac{\exp(-i\pi(\eta+\frac{1+\tau}{2}))}{\exp(-i\frac{\pi(1+\tau)}{2})}\rightarrow 1,\\
J_z\sim \frac{\sin(\pi(\eta+\frac{1}{2}))}{\sin(\frac{\pi}{2})}\rightarrow \cos\pi\eta,
\ea
\end{subequations}
and 
\ba
\alpha(u)\rightarrow \frac{\sin(\pi(u+\eta))}{\sin(\pi\eta)},\quad \beta(u)\rightarrow \frac{\sin(\pi u)}{\sin(\pi\eta)},\quad \gamma(u)\rightarrow 1,\quad \delta(u)\rightarrow 0.
\ea
One can further take the limit $\eta\rightarrow 0$ with the spectral parameter rescaled by $u\rightarrow u\eta$ to go to the XXX limit. 

Let us introduce a new notation, 
\ba
[x]:=\frac{\sin(\pi x)}{\sin(\pi\eta)},
\ea
so that the R-matrix of XXZ spin chain can be expressed as 
\ba
{\bf R}^\text{XXZ}(u)=\lt(\begin{array}{cccc}
[u+\eta] &  &  & \\
 & [u] & [\eta] & \\
 & [\eta] & [u] & \\
 & & & [u+\eta]\\
\end{array}\rt).
\ea
We also note that in the XXX limit, $[u]\rightarrow u$ and $[u+\eta]\rightarrow u+1$. 

In the case of the XXZ spin chain, if we use the notation 
\ba
\iota(\theta)=\lt(\begin{array}{cc}
1 & 0\\
0 & e^{i\theta}\\
\end{array}\rt),
\ea
then we can confirm that $\iota(\theta)\otimes \iota(\theta)$ commutes with the R-matrix, 
\ba
\lt[\iota(\theta)\otimes \iota(\theta),{\bf R}^\text{XXZ}(u)\rt]=0.\label{twist-com}
\ea
This means that the $\theta$-depending transfer matrix 
\ba
\mathsf{t}(u;\theta):={\rm tr}_0\iota_0(\theta){\bf T}_0(u),\label{transfer-twist}
\ea
gives rise to an integrable closed XXZ spin chain with twisted periodic boundary condition, 
\ba
\sigma^{(L+1)}_{x,y,z}=e^{\frac{i}{2}\theta\sigma_z}\sigma^{(1)}_{x,y,z}e^{-\frac{i}{2}\theta\sigma_z}.
\ea
However, we note that the commutation relation (\ref{twist-com}) does not hold for the more general XYZ R-matrix, unless $e^{i\theta}=\pm 1$.

\subsection{Open spin chain}

The open spin chains are more interesting to us in this article. The transfer matrix of an open chain is given by \cite{Sklyanin:1988yz}
\ba
\mathsf{t}(u)={\rm tr}_0K^-_0(u){\bf T}(u)K^+_0(u){\bf T}^{-1}(-u).
\ea
${\bf T}(u)\in {\rm End}(V^{(0)}\otimes V^{\otimes L})$ is usually taken to be the same one as in the closed chain, (\ref{mono-mat}), and $K^\pm(u)\in {\rm End}(V^{(0)})$ stands for the boundary operators, which respectively satisfy the boundary Yang-Baxter equations 
\ba
{\bf R}_{12}(\lambda_1-\lambda_2)K^+_1(\lambda_1){\bf R}_{21}(\lambda_1+\lambda_2)K^+_2(\lambda_2)=K^+_2(\lambda_2){\bf R}_{12}(\lambda_1+\lambda_2)K^+_1(\lambda_1){\bf R}_{21}(\lambda_1-\lambda_2),\label{b-YBE}
\ea
and 
\ba
&&{\bf R}_{12}(-\lambda_1+\lambda_2)K^{-\ t}_1(\lambda_1){\bf R}_{21}(-\lambda_1-\lambda_2-2\eta)K^{-\ t}_2(\lambda_2)\nn\\
 &&=K^{-\ t}_2(\lambda_2){\bf R}_{12}(-\lambda_1-\lambda_2-2\eta)K^{-\ t}_1(\lambda_1){\bf R}_{21}(-\lambda_1+\lambda_2),\label{b-YBE-2}
\ea
with $\eta$ the characteristic parameter of the system, s.t. 
\ba
{\bf R}^{t_1}(u){\bf R}^{t_1}(-u-2\eta)=\rho'(u){\bf I},
\ea
for some function $\rho'(u)$. As has been shown in (\ref{ref-uni}), $\rho'(u)$ for the XYZ R-matrix is given by 
\ba
\rho'(u)=-\frac{\sigma(u+2\eta)\sigma(u)}{\sigma(\eta)^2}.
\ea

Note that we can rewrite the RTT-relation (\ref{RTT}) into 
\ba
{\bf T}_{2}^{-1}(\lambda_2){\bf R}_{12}(\lambda_1-\lambda_2){\bf T}_{1}(\lambda_1)={\bf T}_{1}(\lambda_1){\bf R}_{12}(\lambda_1-\lambda_2){\bf T}^{-1}_{2}(\lambda_2),
\ea
and 
\ba
{\bf T}_{1}^{-1}(\lambda_1){\bf R}_{12}(\lambda_2-\lambda_1){\bf T}_{2}(\lambda_2)={\bf T}_{2}(\lambda_2){\bf R}_{12}(\lambda_2-\lambda_1){\bf T}^{-1}_{1}(\lambda_1),
\ea
therefore 
\ba
\tilde{K}_0^+(u):= {\bf T}_{0}(u)K^+(u){\bf T}^{-1}_{0}(-u),\label{def-K+}
\ea
is also a solution to the boundary Yang-Baxter equation (\ref{b-YBE}), that is to say, we can alternatively express the transfer matrix of the open spin chain as 
\ba
\mathsf{t}(u)={\rm tr}_0K_0^-(u)\tilde{K}_0^+(u).\label{transf-open}
\ea

The commutation relation (\ref{com-pro-transfer}) can also be shown in this case with the following calculation, 
\ba
&&\mathsf{t}(u)\mathsf{t}(u')={\rm tr}_{0,0'}K^-_0(u)K^-_{0'}(u')\tilde{K}^+_0(u)\tilde{K}^+_{0'}(u')={\rm tr}_{0,0'}K^{-\ t}_0(u)K^-_{0'}(u')\tilde{K}^{+\ t}_0(u)\tilde{K}^+_{0'}(u')\nn\\
&&=\rho^{\prime\ -1}(-u-u'-2\eta){\rm tr}_{0,0'}K^{-\ t}_0(u)K^-_{0'}(u'){\bf R}^{t_{0'}}_{00'}(-u-u'-2\eta){\bf R}^{t_0}_{00'}(u+u')\tilde{K}^{+\ t}_0(u)\tilde{K}^+_{0'}(u')\nn\\
&&=\rho^{\prime\ -1}(-u-u'-2\eta){\rm tr}_{0,0'}\lt[\lt(K^{-\ t}_0(u){\bf R}_{00'}(-u-u'-2\eta)K^{-\ t}_{0'}(u')\rt)^{t_{0'}}\rt.\nn\\
&&\qquad\lt.\times \lt(\tilde{K}_0^+(u){\bf R}_{00'}(u+u')\tilde{K}^+_{0'}(u')\rt)^{t_0}\rt]\nn\\
&&=\rho^{\prime\ -1}(-u-u'-2\eta){\rm tr}_{0,0'}\lt[\lt(K^{-\ t}_0(u){\bf R}_{00'}(-u-u'-2\eta)K^{-\ t}_{0'}(u')\rt)^{t_{00'}}\tilde{K}_0^+(u){\bf R}_{00'}(u+u')\tilde{K}^+_{0'}(u')\rt]\nn\\
&&=\rho^{\prime\ -1}(-u-u'-2\eta)\rho^{-1}(u'-u)\nn\\
&&\qquad\times {\rm tr}_{0,0'}\lt[\lt({\bf R}_{00'}(u'-u)K^{-\ t}_0(u){\bf R}_{00'}(-u-u'-2\eta)K^{-\ t}_{0'}(u')\rt)^{t_{00'}}\rt.\nn\\
&&\lt.\times {\bf R}_{00'}(u-u')\tilde{K}_0^+(u){\bf R}_{00'}(u+u')\tilde{K}^+_{0'}(u')\rt]\nn\\
&&=\rho^{\prime\ -1}(-u-u'-2\eta)\rho^{-1}(u'-u)\nn\\
&&\qquad\times {\rm tr}_{0,0'}\lt[\lt(K^{-\ t}_{0'}(u'){\bf R}_{00'}(-u-u'-2\eta)K^{-\ t}_0(u){\bf R}_{00'}(u'-u)\rt)^{t_{00'}}\rt.\nn\\
&&\qquad\lt.\times \tilde{K}^+_{0'}(u'){\bf R}_{00'}(u+u')\tilde{K}_0^+(u){\bf R}_{00'}(u-u')\rt]\nn\\
&&=\mathsf{t}(u')\mathsf{t}(u),
\ea
where $t_{00'}=t_0t_{0'}$ the transpose in both of the $0$-th and the $0'$-th auxiliary spaces, and we used the unitarity of the R-matrix in the derivation. 

We focus on the case of diagonal boundary operator, 
\ba
K(u)=\lt(\begin{array}{cc}
e(u) & 0\\
0 & f(u)\\
\end{array}\rt),
\ea
in this article. 
The boundary Yang-Baxter equation, (\ref{b-YBE}), for this diagonal ansatz reduces to
\begin{subequations}
\ba
\alpha(u-v)\delta(u+v)\lt(e(v)f(u)-e(u)f(v)\rt)+\alpha(u+v)\delta(u-v)\lt(e(u)e(v)-f(u)f(v)\rt)=0,\\
\beta(u+v)\gamma(u-v)\lt(e(v)f(u)-e(u)f(v)\rt)+\beta(u-v)\gamma(u+v)\lt(e(u)e(v)-f(u)f(v)\rt)=0,
\ea
\end{subequations} 
and one can further simplify them to one single equation, 
\ba
&&\theta_{0,1/2}(u-v,2\tau)\theta_{1/2,1/2}(u+v,2\tau)\lt(e(v)f(u)-e(u)f(v)\rt)\nn\\
&&+\theta_{0,1/2}(u+v,2\tau)\theta_{1/2,1/2}(u-v,2\tau)\lt(e(u)e(v)-f(u)f(v)\rt)=0.
\ea
Two trivial solutions are $e(u)=\pm f(u)$, but we would like to consider a more non-trivial one. 
By using the identity
\ba
&&\theta_{1/2,1/2}(u\pm v,\tau)\theta_{0,1/2}(u\mp v,\tau)\theta_{0,0}(0,\tau)\theta_{1/2,0}(0,\tau)\nn\\
&&=\theta_{1/2,1/2}(u,\tau)\theta_{0,1/2}(u,\tau)\theta_{0,0}(v,\tau)\theta_{1/2,0}(v,\tau)\pm \theta_{1/2,1/2}(v,\tau)\theta_{0,1/2}(v,\tau)\theta_{0,0}(u,\tau)\theta_{1/2,0}(u,\tau),\nn\\
\ea
we found a solution to the boundary Yang-Baxter equation, 
\ba
e(u)=\frac{\theta_{1/2,1/2}(u+ \xi,2\tau)\theta_{0,1/2}(u- \xi,2\tau)}{\theta_{1/2,1/2}(\xi,2\tau)\theta_{0,1/2}(\xi,2\tau)},\quad f(u)=-\frac{\theta_{1/2,1/2}(u- \xi,2\tau)\theta_{0,1/2}(u+ \xi,2\tau)}{\theta_{1/2,1/2}(\xi,2\tau)\theta_{0,1/2}(\xi,2\tau)},\label{ef-sol}
\ea
where we normalized the boundary operator $K(u)$ s.t. $K(0)={\bf I}$, as we can also see that the R-matrix (\ref{R-XYZ}) trivializes at the same value of $u=0$. In the XXZ limit, the boundary operator becomes 
\ba
K^\text{XXZ}(u)=\frac{1}{[\xi]}\lt(\begin{array}{cc}
[u+\xi] & \\
 & -[u-\xi]\\
 \end{array}\rt).
 \ea

\paragraph{Remark:} One can easily derive the following identities from (\ref{theta-prod-1}) to (\ref{theta-prod-4}), 
\begin{subequations}
\ba
\theta_{0,\frac{1}{2}}(2u,2\tau)=\lt(\prod_{m=1}^\infty \frac{1+q^{2m}}{1-q^{2m}}\rt)\theta_{0,0}(u,\tau)\theta_{0,\frac{1}{2}}(u,\tau),\\
\theta_{\frac{1}{2},\frac{1}{2}}(2u,2\tau)=\lt(\prod_{m=1}^\infty \frac{1+q^{2m}}{1-q^{2m}}\rt)\theta_{\frac{1}{2},\frac{1}{2}}(u,\tau)\theta_{\frac{1}{2},0}(u,\tau).
\ea
\end{subequations}
Then we can rewrite
\begin{subequations}
\ba
e(u)=\frac{\theta_{1/2,1/2}(\frac{u+ \xi}{2},\tau)\theta_{1/2,0}(\frac{u+ \xi}{2},\tau)\theta_{0,1/2}(\frac{u- \xi}{2},\tau)\theta_{0,0}(\frac{u- \xi}{2},\tau)}{\theta_{1/2,1/2}(\xi/2,\tau)\theta_{1/2,0}(\xi/2,\tau)\theta_{0,1/2}(\xi/2,\tau)\theta_{0,0}(\xi/2,\tau)},\\
f(u)=-\frac{\theta_{1/2,1/2}(\frac{u- \xi}{2},\tau)\theta_{1/2,0}(\frac{u- \xi}{2},\tau)\theta_{0,1/2}(\frac{u+ \xi}{2},\tau)\theta_{0,0}(\frac{u+ \xi}{2},\tau)}{\theta_{1/2,1/2}(\xi/2,\tau)\theta_{1/2,0}(\xi/2,\tau)\theta_{0,1/2}(\xi/2,\tau)\theta_{0,0}(\xi/2,\tau)},
\ea
\end{subequations}
and by further using the identity 
\ba
&&\theta_{1/2,1/2}(x\pm y,\tau)\theta_{1/2,0}(x\mp y,\tau)\theta_{0,0}(0,\tau)\theta_{0,1/2}(0,\tau)=\nn\\
&&\theta_{1/2,1/2}(x,\tau)\theta_{1/2,0}(x,\tau)\theta_{0,0}(y,\tau)\theta_{0,1/2}(y,\tau)\pm \theta_{1/2,1/2}(y,\tau)\theta_{1/2,0}(y,\tau)\theta_{0,0}(x,\tau)\theta_{0,1/2}(x,\tau),\nn\\
\ea
we have 
\ba
e(u)=\frac{\theta_{1/2,1/2}(u,\tau)\theta_{1/2,0}(\xi,\tau)+\theta_{1/2,1/2}(\xi,\tau)\theta_{1/2,0}(u,\tau)}{2\theta_{1/2,1/2}(\xi/2,\tau)\theta_{1/2,0}(\xi/2,\tau)\theta_{0,1/2}(\xi/2,\tau)\theta_{0,0}(\xi/2,\tau)}\theta_{0,0}(0,\tau)\theta_{0,1/2}(0,\tau)\nn\\
=\frac{\theta_{0,0}(0,\tau)\theta_{0,1/2}(0,\tau)\theta_{1/2,1/2}(u,\tau)\theta_{1/2,0}(u,\tau)}{2\theta_{1/2,1/2}(\xi/2,\tau)\theta_{1/2,0}(\xi/2,\tau)\theta_{0,1/2}(\xi/2,\tau)\theta_{0,0}(\xi/2,\tau)}\lt(\frac{\theta_{1/2,0}(\xi,\tau)}{\theta_{1/2,0}(u,\tau)}+\frac{\theta_{1/2,1/2}(\xi,\tau)}{\theta_{1/2,1/2}(u,\tau)}\rt),\nn\\
f(u)=\frac{-\theta_{1/2,1/2}(u,\tau)\theta_{1/2,0}(\xi,\tau)+\theta_{1/2,1/2}(\xi,\tau)\theta_{1/2,0}(u,\tau)}{2\theta_{1/2,1/2}(\xi/2,\tau)\theta_{1/2,0}(\xi/2,\tau)\theta_{0,1/2}(\xi/2,\tau)\theta_{0,0}(\xi/2,\tau)}\theta_{0,0}(0,\tau)\theta_{0,1/2}(0,\tau)\nn\\
=\frac{\theta_{0,0}(0,\tau)\theta_{0,1/2}(0,\tau)\theta_{1/2,1/2}(u,\tau)\theta_{1/2,0}(u,\tau)}{2\theta_{1/2,1/2}(\xi/2,\tau)\theta_{1/2,0}(\xi/2,\tau)\theta_{0,1/2}(\xi/2,\tau)\theta_{0,0}(\xi/2,\tau)}\lt(-\frac{\theta_{1/2,0}(\xi,\tau)}{\theta_{1/2,0}(u,\tau)}+\frac{\theta_{1/2,1/2}(\xi,\tau)}{\theta_{1/2,1/2}(u,\tau)}\rt).
\ea
That is to say, one can decompose the boundary operator $K(u)$ as 
\ba
K(u)=\frac{\theta_{0,0}(0,\tau)\theta_{0,1/2}(0,\tau)\theta_{1/2,1/2}(u,\tau)\theta_{1/2,0}(u,\tau)}{2\theta_{1/2,1/2}(\xi/2,\tau)\theta_{1/2,0}(\xi/2,\tau)\theta_{0,1/2}(\xi/2,\tau)\theta_{0,0}(\xi/2,\tau)}\lt(\frac{\theta_{1/2,1/2}(\xi,\tau)}{\theta_{1/2,1/2}(u,\tau)}{\bf I}+\frac{\theta_{1/2,0}(\xi,\tau)}{\theta_{1/2,0}(u,\tau)}\sigma_z\rt).
\ea
We remark that up to the overall scaling (which is a free choice of the boundary operator), the above decomposition matches with that given in \cite{Hou_1995,Fan:1996jq} as a special diagonal case. $\Box$

For (\ref{b-YBE-2}), we again use the diagonal ansatz $\tilde{K}={\rm diag}\lt(e'(u),f'(u)\rt)$, and obtain the following equation, 
\ba
&&\theta_{0,1/2}(-u+v,2\tau)\theta_{1/2,1/2}(-u-v-2\eta,2\tau)\lt(e'(v)f'(u)-e'(u)f'(v)\rt)\nn\\
&&+\theta_{0,1/2}(-u-v-2\eta,2\tau)\theta_{1/2,1/2}(-u+v,2\tau)\lt(e'(u)e'(v)-f'(u)f'(v)\rt)=0.
\ea
We see that by replacing $u\rightarrow -u-\eta$, $v\rightarrow -v-\eta$ in (\ref{ef-sol}), we obtain the following dual solution for the boundary operator, 
\begin{subequations}
\ba
e'(u)=\frac{\theta_{1/2,1/2}(-u-\eta+ \tilde{\xi},2\tau)\theta_{0,1/2}(-u-\eta- \tilde{\xi},2\tau)}{\theta_{1/2,1/2}(\tilde{\xi},2\tau)\theta_{0,1/2}(\tilde{\xi},2\tau)},\\
 f'(u)=-\frac{\theta_{1/2,1/2}(-u-\eta- \tilde{\xi},2\tau)\theta_{0,1/2}(-u-\eta+ \tilde{\xi},2\tau)}{\theta_{1/2,1/2}(\tilde{\xi},2\tau)\theta_{0,1/2}(\tilde{\xi},2\tau)}.\label{ef-dual-sol}
\ea
\end{subequations}

The Bethe ansatz provides a convenient approach to diagonalize the integrable system. In this article, we compare the vacuum equation of the gauge theory with the Bethe ansatz equation of the spin chain. Let us give a brief description on the (algebraic) Bethe ansatz. We can schematically express the transfer matrix as a trace over the auxiliary space of a matrix, 
\ba
\mathsf{t}(u)=:{\rm tr}_0\lt(\begin{array}{cc}
\cA(u) & \cB(u)\\
\cC(u) & \cD(u)\\
\end{array}\rt)=\cA(u)+\cD(u),
\ea
with the entries $\cA(u)$, $\cB(u)$, $\cC(u)$, $\cD(u)$ elements in ${\rm End}(V^{\otimes L})$. For a given ground state $\ket{\Omega}$ of the system, the Bethe ansatz assumes that all the eigenstates of the system takes the form 
\ba
\prod_{i=1}^M\cB(u_i)\ket{\Omega},
\ea
where the set of $\{u_i\}$ is determined by the so-called Bethe ansatz equation (BAE).

We give a more detailed review on how to derive the above Bethe ansatz equation in Appendix~\ref{a:bethe-open}. 

The Hamiltonian of the open chain can be found in a similar way as in (\ref{d-Ham}), and we only write down the Hamiltonian of the open XXZ spin chain with diagonal boundary conditions here \cite{Sklyanin:1988yz}: 
\ba
{\cal H}=\sum_{i=1}^{L-1}{\cal H}_{i,i+1}+\frac{\pi}{2}\cot(\pi\xi_-)\sigma_z^{(1)}+\frac{1}{2}\tan(\pi\eta)\cot(\pi\xi_+)\sigma_z^{(L)},
\label{boundary_Ham}
\ea
where ${\cal H}_{i,i+1}=\frac{1}{2}\lt(\sigma_x^{(i)}\sigma_x^{(i+1)}+\sigma_y^{(i)}\sigma_y^{(i+1)}+\cos(\pi\eta)\sigma_z^{(i)}\sigma_z^{(i+1)}\rt)$ is the building block of the XXZ spin chain, and we omitted some constant terms. We remark that $\xi=0$ forces the corresponding $\sigma_z$ at the boundary to be zero, which can be thought as a fixed-end (or Direchlet) boundary condition, while taking $\xi\rightarrow i\infty$ gives $\cot\xi\rightarrow -i$ and is also a special limit in the spin chain (that minimizes the boundary coupling on the imaginary axis\footnote{In the context of XXZ spin chain, it is very often to take $u$, $\eta$ and $\xi_\pm$ to be pure imaginary.}). 

\section{Effective Twisted Superpotential of 3d $\cN=2$ Theory and 2d $\cN=(2,2)$ Theory}\label{s:gauge-th}

In this section, we quote the expression of the disk partition function of 3d $\cN=2$ theory (on $D^2\times S^1$) and 2d $\cN=(2,2)$ theory (on $D^2$), and then we compute the effective twisted superpotential of the gauge theories. 

\subsection{3d $\cN=2$ theory}

The partition function of 3d $\cN=2$ theory on $D^2 \times S^1$ was computed in \cite{Yoshida:2014ssa}, where the geometry of $D^2\times S^1$ is parameterized as 
\ba
{\rm d}s^2=\ell^2({\rm d}\theta^2+r^2\sin^2\theta{\rm d}\varphi^2)+{\rm d}\tau^2,
\ea
where the $S^1$ circle has a periodicity $\beta\ell$. 

The index on $D^2\times S^1$ is given by the following integral, 
\ba
{\cal I}=\frac{1}{\mid W_G\mid}\int \frac{{\rm d}^N \sigma}{(2\pi)^N}e^{-S_\text{cl}}Z_\text{vec}Z_\text{chi}Z_\text{bd},
\ea
where we denote the Weyl group of $G$ by $W_G$, and the one-loop determinant of the vector multiplet is given by 
\ba
Z_\text{vec}=\prod_{\alpha \in \hat\Delta} e^{\frac{1}{8\beta_2}(\alpha\cdot \sigma)^2}\lt(e^{i\alpha\cdot \sigma};q^2\rt)_\infty,
\ea
with the set of the roots of $G$ denoted by $\hat\Delta$, and the contribution from the chiral multiplet with Neumann boundary condition reads  
\ba
Z^\text{Neu}_\text{chi}=\prod_{w \in R} e^{{\cal E}(iw\cdot \sigma+\Delta \beta_2+im)}\lt(e^{-iw\cdot\sigma-im}q^\Delta;q^2\rt)^{-1}_\infty,
\qquad
q=e^{-\beta_2},
\ea
with the set of the weights of the corresponding representation denoted by $R$, the R-charge of the scalar in the chiral multiplet $\Delta$, and
\ba
{\cal E}(x)=\frac{\beta_2}{12}-\frac{1}{4}x+\frac{1}{8\beta_2}x^2.
\ea
$\beta_1$ is the fugacity of the rotation along $S^1$, $\beta_2$ is the U(1)$_R$ charge fugacity, $\beta \ell=(\beta_1+\beta_2)\ell$ is the circumference of $S^1$. 
The one-loop contribution of chiral multiplet with Direchlet boundary condition reads 
\ba
Z^\text{Dir}_\text{chi}=\prod_{w \in R}e^{-{\cal E}(-iw\cdot\sigma +(2-\Delta)\beta_2-im)}\lt(e^{iw\cdot\sigma+im}q^{2-\Delta};q^2\rt)_\infty,
\ea
and one can confirm that the difference between the chiral multiplet in Direchlet boundary condition and that in Neumann condition is given by a 2d Fermi multiplet living on the boundary, $T^2 = \partial(D^2 \times S^1)$,
\ba
Z^\text{Dir}_\text{chi}=Z_\text{2d\ Fermi}Z^\text{Neu}_\text{chi},
\ea
with 
\ba
Z_\text{2d\ Fermi}=\prod_{w \in R} e^{-2{\cal E}(iw\cdot\sigma+\Delta\beta_2+F_lM_l)}\theta(e^{-iw\cdot\sigma-F_lM_l}q^{\Delta};q^2),
\ea
where the $\theta$-function here is defined as 
\ba
\theta(y;q)=\prod_{n=0}^\infty (1-yq^n)(1-y^{-1}q^{n+1})=\frac{1}{(1-y^{-1})}(y;q)_\infty (y^{-1};q)_\infty, 
\ea
which is equivalent to $\theta_{1/2,1/2}$ up to the variable change (See~\eqref{theta-prod-4}).

The classical piece depends on the FI-term and the boundary Chern-Simons term (defined on the boundary $T^2=S^1\times S^1$), $S_\text{cl}=S_\text{FI}-S_\text{bCS}$, 
\ba
-S_\text{FI}=2\pi i\ell\zeta {\rm tr}\sigma,
\ea
and 
\ba
S_\text{bCS}=\frac{\kappa}{4\beta}{\rm tr}\sigma^2.
\ea

\paragraph{Remark:} As noted in \cite{Yoshida:2014ssa}, when we focus on the special case of 3d $\cN=4$ theory, the $D^2\times S^1$ partition function can be identified with the $\Omega$-background partition function $\mathbb{C}_{q^2}\times S^1$ presented in \cite{Aganagic:2013tta} (with proper boundary conditions on $D^2$ chosen).\footnote{They basically discuss 3d $\cN=2^*$ theory in \cite{Aganagic:2013tta}, which is the mass deformation of $\cN=4$ with the adjoint matter. In fact, they start with 5d $\cN = 1$, then discuss 3d theory by considering the Higgs branch locus.} The $\cN=4$ hypermultiplet is decomposed into an $\cN=2$ chiral multiplet in fundamental representation with Neumann b.c. and an $\cN=2$ chiral multiplet in anti-fundamental representation with Direchlet b.c. imposed. The $\cN=4$ vector multiplet is decomposed into an $\cN=2$ vector multiplet and an $\cN=2$ chiral multiplet in the adjoint representation with Neumann boundary condition. Under this identification, $q^2$ is mapped to the $\Omega$-background parameter, and the R-charge fugacity $q^\Delta$ is identified with $q/t$, where $t^2$ is the fugacity parameter of the vector U(1)$_{F=J_L+J_R}$ symmetry in the SU(2)$_L\times$SU(2)$_R$ R-symmetry of 3d $\cN=4$ SUSY algebra ($t$ can also be understood as the adjoint mass). We can therefore rescale $\beta_2\rightarrow \beta_2\epsilon/2$ (together with $\Delta=1-\frac{2}{\epsilon}\log_qt$)  in the partition function, and take $\epsilon\rightarrow 1$ to go to the classical limit. We will perform this procedure in section \ref{s:eff-pot} to compute the effective twisted superpotential of 3d $\cN=2$ theories.

\paragraph{2d $\cN=(2,2)$ theory}

The vortex partition of 2d $\cN=(2,2)$ theory on the plane $\mathbb{C}$ with the $\Omega$-background can be obtained in the zero radius limit of 3d partition function, but it can alternatively be computed as a dimensional reduction of 4d $\cN = 1$ gauge theory on the geometry,%
\footnote{%
See \cite{Kimura:2018axa} for a related discussion.
}
\ba
{\rm d}s^2={\mid {\rm d}z-iz(\epsilon{\rm d}w+\bar{\epsilon}{\rm d}\bar{w})\mid^2}+\mid{\rm d}w\mid^2,
\ea
where $z=x_1+ix_2$, $w=x_3+ix_4$. The explicit expression is known as (See, e.g. \cite{Fujimori:2015zaa})
\ba
{\cal I}_{a,\mu}=\int_{C_{a,\mu}}\lt(\prod_{i=1}^r\frac{{\rm d}\sigma_i}{2\pi i\epsilon}\rt)\exp\lt(-\frac{2\pi i\sigma\cdot\tau}{\epsilon}\rt)\lt(\prod_{\alpha\in\hat\Delta}\Gamma\lt(\frac{\alpha\cdot \sigma}{\epsilon}\rt)\rt)^{-1}\prod_{w\in R}\prod_{a=1}^{N_f}\Gamma\lt(\frac{w\cdot \sigma+m_a}{\epsilon}\rt),\label{part-exp}
\ea
where $\{\tau\}$ are the renormalized FI parameters for all the U(1) gauge groups, and the contour $C_{a,\mu}$ picks up the poles satisfying 
\ba
\mu_i\cdot \sigma=-m_{a_i}-\epsilon k_i,\label{2d-pole}
\ea
with $k_{1,\dots,r}\in\mathbb{Z}^r_{\geq 0}$ and $a_{1,\dots,r}$ picking out $r$ flavor labels. $\mu_i$ is a weight vector specifying the poles we pick. The above partition function can also be viewed as a disk partition function of 2d $\cN=(2,2)$ theory with Neumann boundary condition imposed on chiral multiplets \cite{Sugishita:2013jca,Honda:2013uca,Hori:2013ika}. 

\subsection{Effective potential in 3d and 2d}\label{s:eff-pot}

Let us extract out the effective potential of the gauge theory (in the classical limit of the $\Omega$-background, $\epsilon\rightarrow 0$). One has 
\ba
{\cal I}\sim \exp\lt(\frac{1}{\epsilon}W_\text{eff}(\sigma^\ast,m)\rt).
\ea

To compute the effective potential in 3d, we need the formula 
\ba
\epsilon\log(e^{ix};q^2)_\infty=\epsilon\sum_{n=0}^\infty\log(1-e^{ix}q^{2n})=-\epsilon\sum_{n=0}^\infty\sum_{m=1}^\infty \frac{1}{m}e^{imx}q^{2mn}\nn\\
=-\epsilon \sum_{m=1}^\infty \frac{1}{m}e^{imx}\frac{1}{1-e^{-m\beta_2\epsilon}}\xrightarrow{\epsilon \to 0} -\frac{1}{\beta_2}\sum_{m=1}^\infty \frac{1}{m^2}e^{imx}=-\frac{1}{\beta_2}\operatorname{Li}_2(e^{ix}),
\ea
and a related useful identity is given by 
\ba
\exp\lt(\frac{\partial}{\partial x}\operatorname{Li}_2(e^{\pm x})\rt)=(1-e^{\pm x})^{\mp 1}.\label{id-diff}
\ea

Under the rescaling $\beta_2\rightarrow \beta_2\epsilon/2$ and $\Delta\rightarrow 1+\frac{2i}{\epsilon}\tilde{c}$, we obtain 
\ba
W^\text{3d}_\text{eff}(\sigma,m)=-\frac{1}{\beta_2}\sum_{\alpha \in \hat\Delta} \operatorname{Li}_2(e^{i\alpha\cdot \sigma})+\frac{1}{4\beta_2}\sum_{\alpha \in \hat\Delta} (\alpha\cdot\sigma)^2+\frac{1}{\beta_2}\sum_{w \in R} \sum_{a=1}^{N_f}\operatorname{Li}_2(e^{-iw\cdot \sigma-im_a-i\beta_2\tilde{c}})\nn\\
-\frac{1}{4\beta_2}\sum_{w \in R} \sum_{a=1}^{N_f}(w\cdot\sigma+m_a+\beta_2\tilde{c})^2+2\pi i\ell\zeta{\rm tr}\sigma,\label{3d-eff}
\ea
where all the chiral multiplets are put to satisfy the Neumann boundary condition, and we also rescaled the FI-term $\zeta\rightarrow \zeta/\epsilon$. 
To switch the $a$-th chiral multiplet to the Direchlet boundary condition, we simply need to add a contribution 
\ba
W^\text{Fermi}(\sigma,m_a)=\sum_{w \in R}\frac{1}{2\beta_2}(w\cdot\sigma+m_a+\beta_2\tilde{c})^2-\frac{1}{\beta_2}\sum_{w \in R} \lt(\operatorname{Li}_2(e^{iw\cdot\sigma+im_a+i\beta_2\tilde{c}})+\operatorname{Li}_2(e^{-iw\cdot\sigma-im_a-i\beta_2\tilde{c}})\rt)\nn\\
=\frac{\pi^2}{3\beta_2}-\frac{\pi}{\beta_2}\sum_{w \in R} (w\cdot\sigma+m_a+\beta_2\tilde{c}),\nn\\
\ea
where we used the following identity, 
\ba
\operatorname{Li}_2(e^x)+\operatorname{Li}_2(e^{-x})=\frac{\pi^2}{3}-i\pi x-\frac{x^2}{2},\quad x/i\in[0,2\pi).
\ea
We remark that as discussed in the case of 2d $\cN=(2,2)$ theory, (\ref{2d-pole}), the poles picked up in the contour integral take the form 
\ba
\mu_i\cdot \sigma=-m_{a_i}-\epsilon \beta_2 k_i-\epsilon\beta_2 \Delta,
\ea
for some weight vector $\mu_i$ again. In the $\epsilon\rightarrow 0$ limit, the contour integral in the partition function simply forces $\sigma$ to take a specific value $\sigma^\ast_a$ as a linear function of $m_a$'s. That is why we are allowed to treat the effective potential as a normal function of the variables $\sigma_a=\sigma^*_a$, the on-shell values of the critical configuration. 

In the case of 2d theory, we can take the log of the integrand of (\ref{part-exp}) by using Stirling's formula, 
\ba
\Gamma(z+1)\sim \sqrt{2\pi z}(z/e)^z, 
\ea
to obtain 
\ba
W_\text{eff}(\sigma,m)=-\sum_{\alpha\in\hat\Delta} \alpha\cdot \sigma(\log\alpha\cdot\sigma-1)+\sum_{w\in R}\sum_{a=1}^{N_f}(w\cdot\sigma+m_a)\lt(\log(w\cdot\sigma+m_a)-1\rt)\nn\\
-\sum_{w\in R}\sum_{a=1}^{N_f}(w\cdot\sigma+m_a)\log\epsilon-2\pi i \tau\cdot \sigma.\label{pre-eff-W}
\ea
Note that 
\ba
-\sum_{\alpha\in\hat\Delta} \alpha\cdot \sigma(\log\alpha\cdot\sigma-1)=-\sum_{\alpha\in\hat\Delta_+}\lt(\alpha\cdot \sigma\log\alpha\cdot\sigma-\alpha\cdot \sigma\log\lt((-\alpha)\cdot\sigma\rt)\rt)\nn\\
=\sum_{\alpha\in\hat\Delta_+}\alpha\cdot\sigma\log(-1)=2\pi i\sum_{\alpha\in\hat\Delta_+}\frac{\alpha}{2}\cdot \sigma=2\pi i\rho\cdot \sigma,
\ea
with the Weyl vector given by a half sum of the positive roots (their set denoted by $\hat\Delta_+$), $\rho = \frac{1}{2} \sum_{\alpha \in \hat\Delta_+} \alpha$. 
For theories we consider in this article (those have the same number of fundamental and anti-fundamental matters), $\sum_{w\in R}\sum_{a=1}^{N_f}(w\cdot\sigma+m_a)$ vanishes (up to a constant term), and we remark that such restriction to special matter contents agrees with the cancellation condition of U(1)$_R$ anomaly in 2d. We finally arrive at 
\ba
W_\text{eff}(\sigma,m)=-2\pi i \tau\cdot \sigma+2\pi i\rho\cdot\sigma +\sum_{w\in R}\sum_{a=1}^{N_f}(w\cdot\sigma+m_a)\lt(\log(w\cdot\sigma+m_a)-1\rt).\label{eff-W-2d}
\ea


One can alternatively take a 2d limit of the 3d effective potential (\ref{3d-eff}) to derive the the effective potential of 2d $\cN=(2,2)$ theories. To see this, we rescale $\sigma$ to $\beta \ell\sigma$, and $m_a$ to $\beta \ell m_a$, then take the $\beta\sim \beta_2\rightarrow 0$ limit. It is very clear that the quadratic terms in (\ref{3d-eff}) vanish in this limit, and thus do not appear in the 2d effective potential. (\ref{eff-W-2d}) is reproduced (up to an overall scale of $\ell$ and irrelevant constant terms) in this limit by using the following formula, 
\ba
\operatorname{Li}_2(e^x)=x\lt(1-\log(-x)\rt)+\sum_{k=0,k\neq 1}^\infty \frac{\zeta(2-k)}{k!}x^k,
\ea
and identifying $i\ell\sigma$ in 3d with $\sigma$ in 2d. 
A similar contribution as that proportional to $\log\epsilon$ in (\ref{pre-eff-W}), 
\ba
\sum_{w\in R}\sum_{a=1}^{N_f}(w\cdot \sigma+m_a)\log \beta \ell,
\ea
also appears in the 2d limit of the 3d effective potential, and we again expect it to sum to a constant term for the same reason as in the 2d case.

On the other hand, 3d $\cN=2$ theories on $D^2\times S^1$ can be uplift to 4d $\cN=1$ theories on $D^2\times T^2$~\cite{Longhi:2019hdh}, and one would expect the Bethe/Gauge correspondence works parallelly in 4d. However, the computation is more complicated, especially in the case of XYZ open spin chain~\cite{Fan:1996jq}, so we plan to present the details of the correspondence in 4d in a near-future work.

\section{Dictionary between A-type Gauge Theories and Closed Spin Chains}\label{s:dic-A}

Let us now reproduce the known dictionary between the closed spin chains and SU-type gauge theories. First recall that the Bethe ansatz equation in the twisted closed spin chain (derived in Appendix \ref{a:bethe-open}) is given by 
\ba
\prod_{a=1}^L\frac{[u_i+\eta/2+\eta s_a-\vartheta_a]}{[u_i+\eta/2-\eta s_a-\vartheta_a]}=e^{i\theta}\prod_{j\neq i}^m\frac{[u_i-u_j+\eta]}{[u_i-u_j-\eta]}.
\ea

Correspondingly, we consider 3d $\cN=2$ U($N$) theory with one adjoint chiral multiplet, $N_f$ fundamental matters, among which the first $N_d$ obey the Direchlet boundary condition, and $N'_f$ anti-fundamental matters, among which the first $N'_d$ obey the Direchlet boundary condition. The effective potential for this theory is given by 
\ba
W^\text{3d}_\text{eff}(\sigma,m)=-\frac{1}{\beta_2}\sum_{i\neq j}^N \operatorname{Li}_2(e^{i(\sigma_i-\sigma_j)})+\frac{1}{\beta_2}\sum_{i\neq j}^N \operatorname{Li}_2(e^{-i(\sigma_i-\sigma_j)-im_\text{adj}})-\frac{1}{4\beta_2}\sum_{i\neq j}^N (\sigma_i-\sigma_j+m_\text{adj})^2\nn\\
+\frac{1}{\beta_2}\sum_{i=1}^N \sum_{a=1}^{N_f}\operatorname{Li}_2(e^{-i\sigma_i-im_a})-\frac{1}{4\beta_2}\sum_{i=1}^N \sum_{a=1}^{N_f}(\sigma_i+m_a)^2-\frac{1}{\beta_2}\sum_{i=1}^N \sum_{a=1}^{N'_f}\operatorname{Li}_2(e^{i\sigma_i-i\bar{m}_a})\nn\\-\frac{1}{4\beta_2}\sum_{i=1}^N \sum_{a=1}^{N'_f}(\sigma_i-\bar{m}_a)^2+\frac{1}{4\beta_2}\sum_{i\neq j} (\sigma_i-\sigma_j)^2
+2\pi i\ell\zeta{\rm tr}\sigma\nn\\
-\frac{\pi}{\beta_2}\sum_{i=1}^N\sum_{a=1}^{N_d}(\sigma_i+m_a)
-\frac{\pi}{\beta_2}\sum_{i=1}^N\sum_{a=1}^{N'_d}(-\sigma_i+\bar{m}_a),\nn\\
\ea
where several constant terms are dropped in the simplification and we absorbed $\beta_2\tilde{c}$ into the mass parameters. 
The vacuum equation given by
\ba
\exp\lt(\beta_2 i\frac{\partial}{\partial\sigma}W^\text{3d}_\text{eff}(\sigma,m)\rt)=1,
\ea
reads 
\ba
 e^{-\frac{i}{2}\sum_{a=1}^{N_f}(\sigma_i+m_a)-\frac{i}{2}\sum_{a=1}^{N'_f}(\sigma_i-\bar{m}_a)} \prod_{j\neq i}\frac{1-e^{i(\sigma_j-\sigma_i)}}{1-e^{i(\sigma_i-\sigma_j)}} \frac{1-e^{i(\sigma_i-\sigma_j+m_\text{adj})}}{1-e^{i(\sigma_j-\sigma_i+m_\text{adj})}}\frac{\prod_{a=1}^{N'_f}(1-e^{i(\sigma_i-\bar{m}_a)})}{\prod_{a=1}^{N_f}(1-e^{-i(\sigma_i+m_a)})}\nn\\
\times e^{-2\pi \beta_2\ell\zeta-\pi i(N_d-N'_d)}=1,
\ea
where we used (\ref{id-diff}). 
We can simplify it to 
\ba
(-1)^{N'_f+N_d-N'_d}e^{-2\pi \beta_2\ell\zeta}\prod_{j\neq i}\frac{\sin(\sigma_i-\sigma_j+m_\text{adj})}{\sin(\sigma_i-\sigma_j-m_\text{adj})}\frac{\prod_{a=1}^{N'_f}\sin(\sigma_i-\bar{m}_a)}{\prod_{a=1}^{N_f}\sin(\sigma_i+m_a)}=1.
\ea
We note that the adjoint matter is important to cancel the factor $e^{i(\sigma_i-\sigma_j)}$ from the vector multiplet. 

In the case $N_f=N'_f$ ($\cN=2^\ast$ theories), one can identify the above vacuum equation with the Bethe ansatz equation of closed spin chain with $L$ sites and $m$ Bethe roots. The dictionary is given by 
\begin{subequations}
\ba
&m\leftrightarrow N,\quad L\leftrightarrow N_f,&\\
&\pi u_i\leftrightarrow \sigma_i,&\\
&\pi \eta\leftrightarrow m_\text{adj},&\\
&i\theta \leftrightarrow \pi i(N_f+N_d+N'_d)-2\pi \beta_2\ell\zeta.&\label{theta-map}
\ea
\end{subequations}
We also see that we need to specify the mass parameter of chiral multiplets to 
\ba
\pi (\eta/2+\eta s_a-\vartheta_a)\leftrightarrow m_a,\quad \pi (\eta/2-\eta s_a-\vartheta_a)\leftrightarrow -\bar{m}_a.
\ea

The correspondence here certainly works in parallel after taking the 2d limit $\beta\sim\beta_2\rightarrow 0$ in the gauge theory and the XXX limit, $[u]\rightarrow u$, of the spin chain.

\section{Vacuum Equations and Bathe Ansatz Equations of Open Spin Chain}\label{s:corr-open}

In this section, we explore the Bethe/Gauge correspondence between gauge theories with SO and Sp gauge groups and open spin chains with diagonal boundary conditions. There are mainly two reasons for which we consider the open spin chains as the dual integrable system, instead of the closed spin chain. Firstly the symmetry $\sigma\leftrightarrow -\sigma$ in the effective potential of SO and Sp gauge theories naturally appears in the Bethe ansatz equation of open spin chains with diagonal boundary conditions. Secondly, as we can see from the definition of the open chain transfer matrix, (\ref{transf-open}) and (\ref{def-K+}), it can be viewed as a ``folded" version of the closed chain transfer matrix, and this ``folding" process exactly corresponds to the effect of the orientifold added to the brane construction of SO and Sp gauge theories. We may further interpret the boundary operators $K^\pm$ as the realization of the orientifold in the integrable system. 

Let us recall the effective potential of a general 3d $\cN=2^\ast$ theory (\ref{3d-eff}) with a zero FI-term, 
\ba
W^\text{3d}_\text{eff}(\sigma,m)=-\frac{1}{\beta_2}\sum_{\alpha \in \hat\Delta} \operatorname{Li}_2(e^{i\alpha\cdot \sigma})+\frac{1}{4\beta_2}\sum_{\alpha \in \hat\Delta} (\alpha\cdot\sigma)^2+\frac{1}{\beta_2}\sum_{w \in R} \sum_{a=1}^{N_f}\operatorname{Li}_2(e^{-iw\cdot \sigma-im_a-i\beta_2\tilde{c}})\nn\\
-\frac{1}{4\beta_2}\sum_{w \in R} \sum_{a=1}^{N_f}(w\cdot\sigma+m_a+\beta_2\tilde{c})^2.
\ea
We discuss the connection between the gauge theory results and the open spin chain models based on this expression.

\subsection{SO($2N$) theory}

In the case of an SO($2N$) gauge theory with $N_f$ chiral multiplet in the vector representation, the effective potential is specified to 
\ba
&&W^\text{3d}_\text{eff}(\sigma,m)=-\frac{1}{\beta_2}\sum_{i<j}^N \operatorname{Li}_2(e^{i(\pm \sigma_i\pm \sigma_j)})+\frac{1}{4\beta_2}\sum_{i<j}^N (\pm\sigma_i\pm\sigma_j)^2+\frac{1}{\beta_2}\sum_{i<j}^N \operatorname{Li}_2(e^{-i(\pm\sigma_i\pm\sigma_j)-i\beta_2\tilde{c}})\nn\\
&&-\frac{1}{4\beta_2}\sum_{i<j}^N (\pm\sigma_i\pm\sigma_j+\beta_2\tilde{c})^2+\frac{1}{\beta_2}\sum_{i=1}^N \sum_{a=1}^{N_f}\operatorname{Li}_2(e^{-(\pm i\sigma_i+im_a+i\beta_2\tilde{c})})-\frac{1}{4\beta_2}\sum_{i=1}^N \sum_{a=1}^{N_f}(\pm\sigma_i+m_a+\beta_2\tilde{c})^2,\nn\\
\ea
as the positive roots of SO($2N$) is given by $e_i\pm e_j$ ($1\leq i<j\leq N$), where $\pm$ stands for a summation over both signs. We did not impose a specific flavor symmetry on the chiral multiplets at the moment. The vacuum equation reads 
\ba
e^{-i\sigma_i}\prod_{j\neq i}\frac{1-e^{-i(\sigma_i\pm \sigma_j)}}{1-e^{i(\sigma_i\pm \sigma_j)}}\frac{1-e^{i(\sigma_i\pm \sigma_j)-i\beta_2\tilde{c}}}{1-e^{-i(\sigma_i\pm \sigma_j)-i\beta_2\tilde{c}}}\prod_{a=1}^{N_f}\frac{1-e^{i\sigma_i-im_a-i\beta_2\tilde{c}}}{1-e^{-i\sigma_i-im_a-i\beta_2\tilde{c}}}=1,
\ea
or 
\ba
\prod_{j\neq i}\frac{\sin(\sigma_i\pm \sigma_j-\beta_2\tilde{c})}{\sin(-\sigma_i\pm \sigma_j-\beta_2\tilde{c})}\prod_{a=1}^{N_f}\frac{\sin(\sigma_i-m_a-\beta_2\tilde{c})}{\sin(-\sigma_i-m_a-\beta_2\tilde{c})}=1.
\ea
Changing the boundary condition of chiral multiplets does not affect the vacuum equation in this case. Compared to the general Bethe ansatz equation for the open spin chain with diagonal boundary conditions, 
\ba
\frac{\lt[u_i+\xi_+-\frac{\eta}{2}\rt]\lt[u_i-\frac{\eta}{2}+\xi_-\rt]\delta_+(u_i)\delta_-(-u_i)}{\lt[u_i-\xi_++\frac{\eta}{2}\rt]\lt[u_i+\frac{\eta}{2}-\xi_-\rt]\delta_+(-u_i)\delta_-(u_i)}\prod_{j\neq i}^m \frac{[u_j-u_i+\eta][u_i+u_j-\eta]}{[u_j-u_i-\eta][u_j+u_i+\eta]}=1,
\ea
the vacuum equation can be mapped to the above Bethe ansatz equation with the boundary condition chosen as 
\ba
\xi_+=i\infty,\quad \xi_-=i\infty,\label{bd-SO}
\ea
and we also map $m\leftrightarrow N$, $\pi\eta\leftrightarrow \beta_2\tilde{c}$. We remark that $\xi=i\infty$ is also a special choice in the boundary Hamiltonian~\eqref{boundary_Ham}, but it is not the Dirichlet boundary condition. 

Further using the explicit form of $\delta^\pm$, 
\ba
\delta^+(u)=\prod_{a=1}^L[u+\eta/2+\eta s_a-\vartheta_a],\quad \delta^-(u)=\prod_{a=1}^L[u+\eta/2-\eta s_a-\vartheta_a],
\ea
we see that $2L\leftrightarrow N_f$ and the mass parameters $\{m_a\}$ have to be paired as 
\ba
\{m_a+\beta_2\tilde{c}\}\leftrightarrow \lt\{-\pi \eta s_a-\frac{\pi\eta}{2}+\pi\vartheta_a,-\pi \eta s_a+\frac{\pi\eta}{2}-\pi\vartheta_a\rt\},
\ea
to establish the duality. We remark that besides the shift by the contribution from the $R$-charge $\tilde{c}$ (correspondingly the part $-\pi\eta s_a$ in the spin chain) the mass parameters are paired in the opposite sign. This is deemed to originate from the Sp($L$) flavor symmetry of the SO-type gauge theory. We remark that this correspondence and dictionary reduce to the map between a 2d gauge theory with SO-type gauge theory and an open XXX spin chain with the same boundary condition as (\ref{bd-SO}). 

\subsection{SO($2N+1$) theory}

In the case of SO($2N+1$) gauge group, all the roots are given by $\{\pm e_i\pm e_j\}$ for all the possible combinations of $i<j$ and $\{\pm e_i\}_{i=1}^N$ (the total number is $2N^2$). The effective potential is thus given by 
\ba
&&W^\text{3d}_\text{eff}(\sigma,m)=-\frac{1}{\beta_2}\sum_{i<j}^N \operatorname{Li}_2(e^{i(\pm \sigma_i\pm \sigma_j)})+\frac{1}{4\beta_2}\sum_{i<j} (\pm\sigma_i\pm\sigma_j)^2+\frac{1}{\beta_2}\sum_{i<j}^N \operatorname{Li}_2(e^{-i(\pm\sigma_i\pm\sigma_j)-i\beta_2\tilde{c}})\nn\\
&&-\frac{1}{4\beta_2}\sum_{i<j}^N (\pm\sigma_i\pm\sigma_j+\beta_2\tilde{c})^2+\frac{1}{\beta_2}\sum_{i=1}^N \sum_{a=1}^{N_f}\operatorname{Li}_2(e^{-(\pm i\sigma_i+im_a+i\beta_2\tilde{c})})-\frac{1}{4\beta_2}\sum_{i=1}^N \sum_{a=1}^{N_f}(\pm\sigma_i+m_a+\beta_2\tilde{c})^2\nn\\
&&-\frac{1}{\beta_2}\sum_{i=1}^N \operatorname{Li}_2(e^{\pm i\sigma_i})+\frac{1}{2\beta_2}\sigma_i^2+\frac{1}{\beta_2}\sum_{i=1}^N \operatorname{Li}_2(e^{\pm i\sigma_i-i\beta_2\tilde{c}})-\frac{1}{4\beta_2}\sum_{i=1}^N(\sigma_i\pm \beta_2\tilde{c})^2.
\ea
The vacuum equation reads 
\ba
e^{-i\sigma_i}\frac{1-e^{-i\sigma_i}}{1-e^{i\sigma_i}}\frac{1-e^{i\sigma_i-i\beta_2\tilde{c}}}{1-e^{-i\sigma_i-i\beta_2\tilde{c}}}\prod_{j\neq i}\frac{1-e^{-i(\sigma_i\pm \sigma_j)}}{1-e^{i(\sigma_i\pm \sigma_j)}}\frac{1-e^{i(\sigma_i\pm \sigma_j)-i\beta_2\tilde{c}}}{1-e^{-i(\sigma_i\pm \sigma_j)-i\beta_2\tilde{c}}}\prod_{a=1}^{N_f}\frac{1-e^{i\sigma_i-im_a-i\beta_2\tilde{c}}}{1-e^{-i\sigma_i-im_a-i\beta_2\tilde{c}}}=1,
\ea
or equivalently 
\ba
\frac{\sin(\sigma_i-\beta_2\tilde{c})}{\sin(\sigma_i+\beta_2\tilde{c})}\prod_{j\neq i}\frac{\sin(\sigma_i\pm \sigma_j-\beta_2\tilde{c})}{\sin(-\sigma_i\pm \sigma_j-\beta_2\tilde{c})}\prod_{a=1}^{N_f}\frac{\sin(\sigma_i-m_a-\beta_2\tilde{c})}{\sin(-\sigma_i-m_a-\beta_2\tilde{c})}=1,
\ea
where we used the fact that the flavor symmetry is expected to be Sp($L$), which means $N_f=2L$ is an even integer. 
The same dictionary maps the above equation to the Bethe ansatz equation of the open spin chain with the boundary condition 
\ba
\xi_+=\frac{\eta}{2},\quad \xi_-=-\frac{\eta}{2}, \quad {\rm for}\ N_f:\ {\rm even},
\ea
which is slightly different from the boundary condition for SO($2N$) theory~\eqref{bd-SO}: the boundary condition parameters are different at two ends of the spin chain.

\subsection{Sp($N$) theory}

Similarly in the case of Sp($N$) gauge theory, all the roots are given by $\{\pm e_i\pm e_j\}$ for $i<j$ and $\{\pm 2e_i\}_{i=1}^N$ (the total number is $2N^2$). The effective potential reads 
\ba
&&W^\text{3d}_\text{eff}(\sigma,m)=-\frac{1}{\beta_2}\sum_{i<j}^N \operatorname{Li}_2(e^{i(\pm \sigma_i\pm \sigma_j)})+\frac{1}{4\beta_2}\sum_{i<j}^N (\pm\sigma_i\pm\sigma_j)^2+\frac{1}{\beta_2}\sum_{i<j}^N \operatorname{Li}_2(e^{-i(\pm\sigma_i\pm\sigma_j)-i\beta_2\tilde{c}})\nn\\
&&-\frac{1}{4\beta_2}\sum_{i<j}^N (\pm\sigma_i\pm\sigma_j+\beta_2\tilde{c})^2+\frac{1}{\beta_2}\sum_{i=1}^N \sum_{a=1}^{N_f}\operatorname{Li}_2(e^{-(\pm i\sigma_i+im_a+i\beta_2\tilde{c})})-\frac{1}{4\beta_2}\sum_{i=1}^N \sum_{a=1}^{N_f}(\pm\sigma_i+m_a+\beta_2\tilde{c})^2\nn\\
&&-\frac{1}{\beta_2}\sum_{i=1}^N \operatorname{Li}_2(e^{\pm 2i\sigma_i})+\frac{2}{\beta_2}\sigma_i^2+\frac{1}{\beta_2}\sum_{i=1}^N \operatorname{Li}_2(e^{\pm 2i\sigma_i-i\beta_2\tilde{c}})-\frac{1}{4\beta_2}\sum_{i=1}^N(2\sigma_i\pm \beta_2\tilde{c})^2,
\ea
and the vacuum equation is given by 
\ba
\frac{\sin^2(2\sigma_i-\beta_2\tilde{c})}{\sin^2(2\sigma_i+\beta_2\tilde{c})}\prod_{j\neq i}\frac{\sin(\sigma_i\pm \sigma_j-\beta_2\tilde{c})}{\sin(-\sigma_i\pm \sigma_j-\beta_2\tilde{c})}\prod_{a=1}^{N_f}\frac{\sin(\sigma_i-m_a-\beta_2\tilde{c})}{\sin(-\sigma_i-m_a-\beta_2\tilde{c})}=1.
\ea
This equation does not directly correspond to a Bethe ansatz equation because of the factor of $2$ in $\sin(2\sigma_i\pm\beta_2\tilde{c})$, but in the 2d limit where $[u]\rightarrow u$, it becomes a Bethe ansatz equation for the open spin chain with boundary condition, 
\ba
\xi_+=0,\quad \xi_-=0.
\ea
As mentioned around \eqref{boundary_Ham}, this corresponds to the Dirichlet boundary condition on the both ends of the spin chain.

We remark that there is a similar factor appeared in the study of $C$-type quiver gauge theories in the literature \cite{Dey:2016qqp,Chen:2018ntf} as the squared factor $\frac{\sin^2(2\sigma_i-\beta_2\tilde{c})}{\sin^2(2\sigma_i+\beta_2\tilde{c})}$ in the above equation.
Such a factor appears in the context of the folding trick to construct the non-simply-laced algebra from the simply-laced algebra.

\section{$A_2$ quiver}\label{s:A2-qui}

In general, the correspondence between gauge theory and spin chain is promoted to the highe rank cases~\cite{Dorey:2011pa,Chen:2011sj,Chen:2012we,Nekrasov:2012xe,Nekrasov:2013xda}.
In this section, we explore the correspondence between $A_2$ quiver, which is the simplest non-trivial quiver gauge theory, and $\mathfrak{sl}_3$ spin chain model.

\subsection{$\mathfrak{sl}_3$ spin chain}

The R-matrix associated to the quantum group $U_q(\widehat{\mathfrak{sl}}_3)$ is known to take the form \cite{PERK1981407}
\ba
{\bf R}(u)=\lt(\begin{array}{ccc|ccc|ccc}
[u+\eta] & & & & & & & & \\
 & [u] & & e^{i\pi u}[\eta] & & & & &\\
 & & [u] & & & & e^{i\pi u}[\eta] & & \\
 \hline
 & e^{-i\pi u}[\eta] & & [u] & & & & &\\
 & & & & [u+\eta] & & & & \\
 & & & & & [u] & & e^{i\pi u}[\eta] & \\
 \hline
 & & e^{-i\pi u}[\eta] & & & & [u] & & \\
 & & & & & e^{-i\pi u}[\eta] & & [u] & \\
 & & & & & & & & [u+\eta]\\
\end{array}
\rt),
\ea
in the convention of this article. 
Needless to say, this R-matrix also satisfies various kinds of basic properties of the R-matrix, and especially 
\ba
{\bf R}(0)={\cal P}.
\ea

The Bethe ansatz of a general periodic spin chain associated to the R-matrix of the Lie algebra $\mathfrak{g}$ is well-known in the literature \cite{Dorey:2006an}. In the case of $\mathfrak{g}=\mathfrak{sl}_3$, there are two sets of Bethe roots, $\{u^{(1)}_i\}_{i=1}^{m_1}$ and $\{u^{(2)}_i\}_{i=1}^{m_2}$. The Bethe ansatz equations are \cite{Babelon:1981un,Babelon:1982gp}
\begin{subequations}
\ba
\prod_{a=1}^{L_1}\frac{[u^{(1)}_i-\vartheta^{(1)}_a+\eta/2]}{[u^{(1)}_i-\vartheta^{(1)}_a-\eta/2]}\prod_{j=1}^{m_2}\frac{[u^{(1)}_i-u^{(2)}_j+\eta/2]}{[u^{(1)}_i-u^{(2)}_j-\eta/2]}=e^{i\delta^{(1)}}\prod_{j=1}^{m_1}\frac{[u^{(1)}_i-u^{(1)}_j+\eta]}{[u^{(1)}_i-u^{(1)}_j-\eta]},\\
\prod_{a=1}^{L_2}\frac{[u^{(2)}_i-\vartheta^{(2)}_a+\eta/2]}{[u^{(2)}_i-\vartheta^{(2)}_a-\eta/2]}\prod_{j=1}^{m_1}\frac{[u^{(2)}_i-u^{(1)}_j+\eta/2]}{[u^{(2)}_i-u^{(1)}_j-\eta/2]}=e^{i\delta^{(2)}}\prod_{j=1}^{m_2}\frac{[u^{(2)}_i-u^{(2)}_j+\eta]}{[u^{(2)}_i-u^{(2)}_j-\eta]}.
\ea
\end{subequations}

\subsection{$A_2$ quiver gauge theory}

Correspondingly, we consider a 3d gauge theory with $A_2$ quiver structure. 
That is we glue two gauge nodes with two bifundamental chiral multiplets (they together form a 3d $\cN=4$ bifundamental hypermultiplet in the massless limit). 
The quiver diagram of the theory we consider is given by 
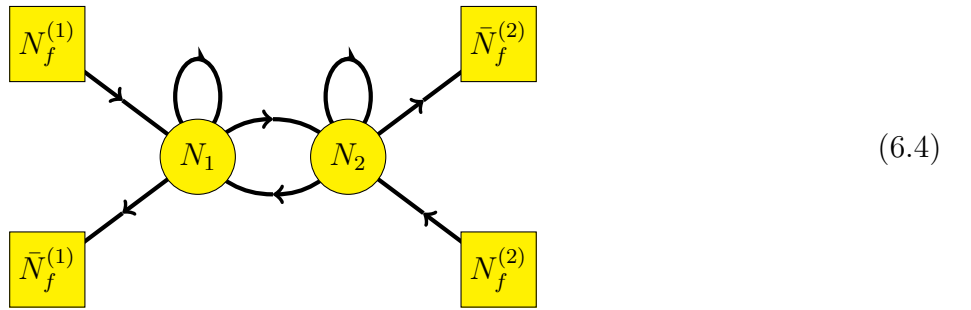
\begin{align}
\begin{tikzpicture}[baseline=(current bounding box.center)]
\draw[ultra thick,->] (-2,1.5)--(-1,0.75);
\draw[ultra thick] (-2,-1.5)--(-1,-0.75);
\draw[ultra thick,<-] (-1,-0.75)--(0,0);
\draw[ultra thick] (-1,0.75)--(0,0);
\draw[ultra thick,->] (0,0) to [out=45,in=180] (1,0.5);
\draw[ultra thick] (1,0.5) to [out=0,in=135] (2,0);
\draw[ultra thick,->] (2,0) to [out=-135,in=0] (1,-0.5);
\draw[ultra thick] (1,-0.5) to [out=180,in=-45] (0,0);
\draw[ultra thick,->] (2,0)--(3,0.75);
\draw[ultra thick] (3,0.75)--(4,1.5);
\draw[ultra thick] (2,0)--(3,-0.75);
\draw[ultra thick,<-] (3,-0.75)--(4,-1.5);
\draw[ultra thick,->-] (0,.3) arc [x radius = .3, y radius = .5, start angle = 270, end angle = -90];
\draw[ultra thick,->-] (2,.3) arc [x radius = .3, y radius = .5, start angle = 270, end angle = -90]; 
\draw[fill=yellow] (0,0) circle (0.5) node {$N_1$};
\draw[fill=yellow] (-2.5,2) rectangle (-1.5,1);
\draw[fill=yellow] (2,0) circle (0.5) node {$N_2$};
\draw[fill=yellow] (3.5,2) rectangle (4.5,1);
\draw[fill=yellow] (-2.5,-2) rectangle (-1.5,-1);
\draw[fill=yellow] (3.5,-2) rectangle (4.5,-1);
 \node at (-2,1.5) {$N_f^{(1)}$};
 \node at (-2,-1.5) {$\bar{N}_f^{(1)}$};
 \node at (4,1.5) {$\bar{N}_f^{(2)}$};
 \node at (4,-1.5) {$N_f^{(2)}$}; 
\end{tikzpicture}
\end{align}
where yellow nodes are used to stand for SU-type gauge groups or flavor symmetry. 
The contribution of these bifundamental matters to the effective potential reads 
\ba
W^\text{3d bfd}_\text{eff}=\frac{1}{\beta_2}\sum_{i=1}^{N_1}\sum_{j=1}^{N_2} \operatorname{Li}_2(e^{-i(\sigma^{(1)}_i-\sigma^{(2)}_j)-im_\text{bfd}})
-\frac{1}{4\beta_2}\sum_{i=1}^{N_1}\sum_{j=1}^{N_2} (\sigma^{(1)}_i-\sigma^{(2)}_j+m_\text{bfd})^2\nn\\
+\frac{1}{\beta_2}\sum_{i=1}^{N_1}\sum_{j=1}^{N_2} \operatorname{Li}_2(e^{-i(\sigma^{(2)}_j-\sigma^{(1)}_i)-im_\text{bfd}})
-\frac{1}{4\beta_2}\sum_{i=1}^{N_1}\sum_{j=1}^{N_2} (\sigma^{(2)}_j-\sigma^{(1)}_i+m_\text{bfd})^2.
\ea

The vacuum equations are then given by 
\begin{subequations}
\ba
e^{i\theta^{(1)}}\prod_{j\neq i}^{N_1}\frac{\sin(\sigma^{(1)}_i-\sigma^{(1)}_j+m^{(1)}_\text{adj})}{\sin(\sigma^{(1)}_i-\sigma^{(1)}_j-m^{(1)}_\text{adj})}\frac{\prod_{a=1}^{\bar{N}^{(1)}_f}\sin(\sigma^{(1)}_i-\bar{m}^{(1)}_a)}{\prod_{a=1}^{N^{(1)}_f}\sin(\sigma^{(1)}_i+m^{(1)}_a)}\prod_{k=1}^{N_2}\frac{\sin(\sigma^{(2)}_k-\sigma^{(1)}_i+m_\text{bfd})}{\sin(\sigma^{(1)}_i-\sigma^{(2)}_k+m_\text{bfd})}=1,\\
e^{i\theta^{(2)}}\prod_{j\neq i}^{N_2}\frac{\sin(\sigma^{(2)}_i-\sigma^{(2)}_j+m^{(2)}_\text{adj})}{\sin(\sigma^{(2)}_i-\sigma^{(2)}_j-m^{(2)}_\text{adj})}\frac{\prod_{a=1}^{\bar{N}^{(2)}_f}\sin(\sigma^{(2)}_i-\bar{m}^{(2)}_a)}{\prod_{a=1}^{N^{(2)}_f}\sin(\sigma^{(2)}_i+m^{(2)}_a)}\prod_{k=1}^{N_1}\frac{\sin(\sigma^{(1)}_k-\sigma^{(2)}_i+m_\text{bfd})}{\sin(\sigma^{(2)}_i-\sigma^{(1)}_k+m_\text{bfd})}=1,
\ea
\end{subequations} 
where $\theta^{(1,2)}$ are expressed in terms of the gauge theory quantities as (\ref{theta-map}). 
If we can adjust the mass parameters to satisfy $m_\text{bfd}=\frac{1}{2}m_\text{adj}^{(1)}=\frac{1}{2}m_\text{adj}^{(2)}\leftrightarrow \frac{\eta}{2}$, then we obtain the $A_2$-type Bethe ansatz equations from the above $A_2$-quiver gauge theory under the map
\begin{subequations}
\ba
\delta^{(i)}\leftrightarrow \theta^{(i)}+i\pi N_i,\quad N^{(i)}_f=\bar{N}^{(i)}_f\leftrightarrow L_i,\quad N_i\leftrightarrow m_i,\\
m^{(i)}_a\leftrightarrow \frac{\eta}{2}-\vartheta^{(i)}_a,\quad \bar{m}^{(i)}_a\leftrightarrow \frac{\eta}{2}+\vartheta^{(i)}_a,
\ea
\end{subequations}
for $i=1,2$. 

\subsection{Open boundary condition}

In the case of open spin chain, the Bethe ansatz has been worked out in \cite{Sun:2017wjh}. We focus on a special diagonal boundary operator of the form 
\ba
K^+(u)=\lt(\begin{array}{ccc}
-e^{i\pi u} [u-\xi] & 0 & 0\\
0 & -e^{i\pi u} [u-\xi] & 0\\
0 & 0 & e^{-i\pi u}[u+\xi]\\
\end{array}\rt),
\ea
and its dual 
\ba
K^-(u)=\lt(\begin{array}{ccc}
e^{-i\pi u+\frac{5}{2}i\pi \eta} [u+\bar{\xi}+3\eta/2] & 0 & 0\\
0 & e^{-i\pi u+\frac{1}{2}i\pi \eta} [u+\bar{\xi}+3\eta/2] & 0\\
0 & 0 & -e^{i\pi u+\frac{3}{2}i\pi \eta}[u-\bar{\xi}+3\eta/2]\\
\end{array}\rt),\nn\\
\ea
in this article. The boundary operator $K^+(u)$ again satisfies (\ref{b-YBE}), while the dual boundary Yang-Baxter equation is slightly modified to 
\ba
&&{\bf R}_{12}(-u_1+u_2)K_1^-(u_1)M^{-1}_1{\bf R}_{21}(-u_1-u_2-3\eta)M_1K^+_2(u_2)\nn\\
&&=K^+_2(u_2)M^{-1}_2{\bf R}_{12}(-u_1-u_2-3\eta)M_2K^+_1(u_1){\bf R}_{21}(u_2-u_1),
\ea
where 
\ba
M={\rm diag}\lt(e^{4\eta},e^{2\eta},1\rt).
\ea
It follows from the crossing unitarity relation, 
\ba
{\bf R}^{t_1}_{12}M_1{\bf R}^{t_1}_{21}(-u-3\eta)M_1^{-1}=\rho''(u){\bf I},
\ea
for some function $\rho''(u)$, in the case of $\mathfrak{sl}_3$ R-matrix. 

The Bethe ansatz equations are 
\begin{subequations}
\ba
\frac{[2u^{(1)}_i-\eta]}{[2u^{(1)}_i+\eta]}\frac{[u^{(1)}_i+\xi+\eta/2][u^{(1)}_i-\bar{\xi}]}{[u^{(1)}_i-\xi-\eta/2][u^{(1)}_i+\bar{\xi}]}\prod_{j=1}^{m_1}\frac{[u^{(1)}_i-u^{(1)}_j+\eta][u^{(1)}_i+u^{(1)}_j+\eta]}{[u^{(1)}_i-u^{(1)}_j-\eta][u^{(1)}_i+u^{(1)}_j-\eta]}\nn\\
\prod_{k=1}^{m_2}\frac{[u^{(1)}_i-u^{(2)}_k-\frac{\eta}{2}][u^{(1)}_i+u^{(2)}_k-\frac{\eta}{2}]}{[u^{(1)}_i-u^{(2)}_k+\frac{\eta}{2}][u^{(1)}_i+u^{(2)}_k+\frac{\eta}{2}]}\prod_{a=1}^{L_1}\frac{[u_i+\theta_a-\frac{\eta}{2}][u_i-\theta_a-\frac{\eta}{2}]}{[u_i+\theta_a+\frac{\eta}{2}][u_i-\theta_a+\frac{\eta}{2}]}=1,\label{bethe-sl3-1}\\
\frac{[2u^{(2)}_i+\eta]}{[2u^{(2)}_i-\eta]}\frac{[u^{(2)}_i+\xi][u^{(2)}_i-\bar{\xi}-\eta/2]}{[u^{(2)}_i-\xi][u^{(2)}_i+\bar{\xi}+\eta/2]}\prod_{j=1}^{m_1}\frac{[u_i^{(2)}-u^{(1)}_j+\eta/2][u^{(2)}_i+u^{(1)}_j+\eta/2]}{[u^{(2)}_i-u^{(1)}_j-\eta/2][u^{(2)}_i+u^{(1)}_j-\eta/2]}\nn\\
\prod_{j=1}^{m_2}\frac{[u^{(2)}_i-u^{(2)}_j-\eta][u^{(2)}_i+u^{(2)}_j-\eta]}{[u^{(2)}_i-u^{(2)}_j+\eta][u^{(2)}_i+u^{(2)}_j+\eta]}=1.\label{bethe-sl3-2}
\ea
\end{subequations}

In the context of gauge theory, there are two possibilities we would like to analyze. One is an Sp($N_1$)-SO($2N_2$) quiver gauge theory, i.e. one gauge node (say the first node) is Sp($N_1$) gauge group and the other (the second node) is SO($2N_2$) gauge group. Since all the representations of Sp and SO Lie algebras are either real or pseudo-real, a 3d $\cN=4$ half-hypermultiplet in these gauge theories is equivalent to one 3d $\cN=2$ chiral multiplet (See~\cite{Hollands:2010xa} for a related discussion in 4d $\cN=2$ theory).
The quiver structure thus looks like the following (SO and Sp nodes are respectively depicted in blue and green). 
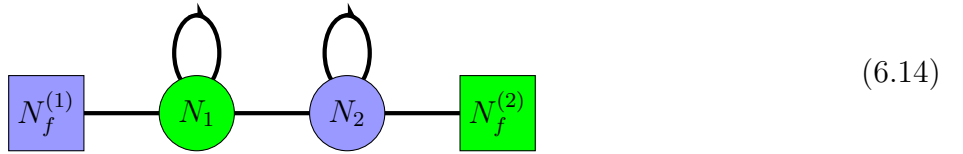
\begin{align}
\begin{tikzpicture}[baseline=(current bounding box.center)]
\draw[ultra thick] (-2,0)--(0,0);
\draw[ultra thick] (2,0)--(3.5,0);
\draw[ultra thick] (0,0)--(2,0);
\draw[ultra thick,->-] (0,.3) arc [x radius = .3, y radius = .5, start angle = 270, end angle = -90];
\draw[ultra thick,->-] (2,.3) arc [x radius = .3, y radius = .5, start angle = 270, end angle = -90];
\draw[fill=green] (0,0) circle (0.5) node {$N_1$};
\draw[fill=blue!40] (-2.5,0.5) rectangle (-1.5,-0.5);
\draw[fill=blue!40] (2,0) circle (0.5) node {$N_2$};
\draw[fill=green] (3.5,0.5) rectangle (4.5,-0.5);
 \node at (-2,0) {$N_f^{(1)}$};
 \node at (4,0) {$N_f^{(2)}$};
\end{tikzpicture}
\end{align}
The effective potential of the bifundamental matter part is given by\footnote{One is allowed to turn on the fugacity parameter $\Delta$ of the 3d $\cN=2$ (bifundamental) chiral multiplet that gives rise to an effective bifundamental mass $m_{bfd}$. In the 2d limit, this corresponds to the twisted mass deformation of the 2d $\cN=(2,2)$ chiral multiplet. } 
\ba
W^\text{3d bfd}_\text{eff}=\frac{1}{\beta_2}\sum_{i=1}^{N_1}\sum_{j=1}^{N_2} \operatorname{Li}_2(e^{-i(\pm \sigma^{(1)}_i+\pm\sigma^{(2)}_j)-im_\text{bfd}})
-\frac{1}{4\beta_2}\sum_{i=1}^{N_1}\sum_{j=1}^{N_2} (\pm\sigma^{(1)}_i+\pm\sigma^{(2)}_j+m_\text{bfd})^2.
\ea
The vacuum equations are found to be
\begin{subequations}
\ba
\frac{\sin^2(2\sigma^{(1)}_i-\beta_2\tilde{c}_1)}{\sin^2(2\sigma^{(1)}_i+\beta_2\tilde{c}_1)}\prod_{j\neq i}^{N_1}\frac{\sin(\sigma^{(1)}_i\pm \sigma^{(1)}_j-\beta_2\tilde{c}_1)}{\sin(-\sigma^{(1)}_i\pm \sigma^{(1)}_j-\beta_2\tilde{c}_1)}\prod_{a=1}^{N^{(1)}_f}\frac{\sin(\sigma^{(1)}_i-m^{(1)}_a-\beta_2\tilde{c}_1)}{\sin(-\sigma^{(1)}_i-m^{(1)}_a-\beta_2\tilde{c}_1)}\nn\\
\times\prod_{j=1}^{N_2}\frac{\sin(\sigma^{(1)}_i\pm\sigma^{(2)}_j-m_\text{bfd})}{\sin(-\sigma^{(1)}_i\pm\sigma^{(2)}_j-m_\text{bfd})}=1,\\
\prod_{j\neq i}\frac{\sin(\sigma^{(2)}_i\pm \sigma^{(2)}_j-\beta_2\tilde{c}_2)}{\sin(-\sigma^{(2)}_i\pm \sigma^{(2)}_j-\beta_2\tilde{c}_2)}\prod_{a=1}^{N^{(2)}_f}\frac{\sin(\sigma^{(2)}_i-m^{(2)}_a-\beta^{(2)}_2\tilde{c}_2)}{\sin(-\sigma^{(2)}_i-m^{(2)}_a-\beta_2\tilde{c}_2)}\prod_{j=1}^{N_1}\frac{\sin(\sigma^{(2)}_i\pm\sigma^{(1)}_j-m_\text{bfd})}{\sin(-\sigma^{(2)}_i\pm\sigma^{(1)}_j-m_\text{bfd})}=1.
\ea
\end{subequations}
We note that the Bethe ansatz equation (\ref{bethe-sl3-1}) and (\ref{bethe-sl3-2}) we want to map to is symmetric about $\sigma^{(1)}\leftrightarrow \sigma^{(2)}$ and $\xi\leftrightarrow \bar{\xi}$. Similar to the case of $A_1$ quiver gauge theory of Sp($N$) gauge group, we again found difficulties to realize the factor $\frac{\sin^2(2\sigma^{(1)}_i-\beta_2\tilde{c}_1)}{\sin^2(2\sigma^{(1)}_i+\beta_2\tilde{c}_1)}$. However, if we take the 2d limit, $\sin\sigma\rightarrow \sigma$, we can choose 
\ba
\xi=0,\quad \bar{\xi}=0,
\ea
to match the Bethe ansatz equations (\ref{bethe-sl3-1}) and (\ref{bethe-sl3-2}). Here we used the dictionary 
\ba
\pi\eta\leftrightarrow \beta_2\tilde{c}_1=\beta_2\tilde{c}_2,\quad -\frac{\pi\eta}{2}\leftrightarrow m_\text{bfd},\label{dict-A2}
\ea
and also added an imaginary site with $\vartheta^{(1)}_0=0$ in the spin chain (that is $L_1-1\leftrightarrow N^{(1)}_f$) to absorb an overall factor of $\frac{[u-\eta/2]^2}{[u+\eta/2]^2}$ in equation (\ref{bethe-sl3-1}). 

Another candidate theory we would like to consider is Sp($N_1$)-SO($2N_2+1$) quiver gauge theory. The bifundamental contribution to the effective potential is 
\ba
W^\text{3d bfd}_\text{eff}=\frac{1}{\beta_2}\sum_{i=1}^{N_1}\sum_{j=1}^{N_2} \operatorname{Li}_2(e^{-i(\pm \sigma^{(1)}_i+\pm\sigma^{(2)}_j)-im_\text{bfd}})
-\frac{1}{4\beta_2}\sum_{i=1}^{N_1}\sum_{j=1}^{N_2} (\pm\sigma^{(1)}_i+\pm\sigma^{(2)}_j+m_\text{bfd})^2\nn\\
+\frac{1}{\beta_2}\sum_{i=1}^{N_1}\operatorname{Li}_2(e^{-i(\pm \sigma^{(1)}_i-im_\text{bfd}})
-\frac{1}{4\beta_2}\sum_{i=1}^{N_1}(\pm\sigma^{(1)}_i+m_\text{bfd})^2
\ea
Then the vacua equations are found to be 
\begin{subequations}
\ba
\frac{\sin^2(2\sigma^{(1)}_i-\beta_2\tilde{c}_1)}{\sin^2(2\sigma^{(1)}_i+\beta_2\tilde{c}_1)}\prod_{j\neq i}^{N_1}\frac{\sin(\sigma^{(1)}_i\pm \sigma^{(1)}_j-\beta_2\tilde{c}_1)}{\sin(-\sigma^{(1)}_i\pm \sigma^{(1)}_j-\beta_2\tilde{c}_1)}\prod_{a=1}^{N^{(1)}_f}\frac{\sin(\sigma^{(1)}_i-m^{(1)}_a-\beta_2\tilde{c}_1)}{\sin(-\sigma^{(1)}_i-m^{(1)}_a-\beta_2\tilde{c}_1)}\nn\\
\times\prod_{j=1}^{N_2}\frac{\sin(\sigma^{(1)}_i\pm\sigma^{(2)}_j-m_\text{bfd})}{\sin(-\sigma^{(1)}_i\pm\sigma^{(2)}_j-m_\text{bfd})}\times\frac{\sin(\sigma^{(1)}_i-m_\text{bfd})}{\sin(-\sigma^{(1)}_i-m_\text{bfd})}=1,\\
\prod_{j\neq i}\frac{\sin(\sigma^{(2)}_i\pm \sigma^{(2)}_j-\beta_2\tilde{c}_2)}{\sin(-\sigma^{(2)}_i\pm \sigma^{(2)}_j-\beta_2\tilde{c}_2)}\prod_{a=1}^{N^{(2)}_f}\frac{\sin(\sigma^{(2)}_i-m^{(2)}_a-\beta^{(2)}_2\tilde{c}_2)}{\sin(-\sigma^{(2)}_i-m^{(2)}_a-\beta_2\tilde{c}_2)}\prod_{j=1}^{N_1}\frac{\sin(\sigma^{(2)}_i\pm\sigma^{(1)}_j-m_\text{bfd})}{\sin(-\sigma^{(2)}_i\pm\sigma^{(1)}_j-m_\text{bfd})}\nn\\
\times\frac{\sin(\sigma^{(2)}_i-\beta_2\tilde{c}_2)}{\sin(\sigma^{(2)}_i+\beta_2\tilde{c}_2)}=1.
\ea
\end{subequations}
Under the same map (\ref{dict-A2}), we note that it is also possible in the 2d limit to choose 
\ba
\xi=0,\quad \bar{\xi}=\frac{\eta}{2},
\ea
to match the vacuum equations with the Bethe ansatz equations. This time we need to further add two imaginary sites with $\vartheta^{(1)}_0=\vartheta^{(1)}_{0'}=0$ ($L_1-2\leftrightarrow N^{(1)}_f$).

\section{$A_r$ quiver}\label{s:Ar_quiver}


The corerspondence between gauge theory and spin chains is promoted to the higher rank cases~\cite{Dorey:2011pa,Chen:2011sj,Chen:2012we,Nekrasov:2012xe,Nekrasov:2013xda}.
From this point of view, $A_r$ quiver gauge theory with SO and Sp symmetry is expected to correspond to $\mathfrak{sl}_{r+1}$ spin chain model with the diagonal-type open boundary condition.
The vacuum equations in such gauge theories are easy to find.  For an Sp gauge node at the $\alpha$-th node, the vacuum equation for $\alpha = 1,3,5,\ldots,2\lfloor\frac{r-1}{2}\rfloor+1$ reads 
\ba
&&\frac{\sin^2(2\sigma^{(\alpha)}_i-\beta_2\tilde{c}_\alpha)}{\sin^2(2\sigma^{(\alpha)}_i+\beta_2\tilde{c}_\alpha)}\prod_{j\neq i}^{N_\alpha}\frac{\sin(\sigma^{(\alpha)}_i\pm \sigma^{(\alpha)}_j-\beta_2\tilde{c}_\alpha)}{\sin(-\sigma^{(\alpha)}_i\pm \sigma^{(\alpha)}_j-\beta_2\tilde{c}_\alpha)}\prod_{j=1}^{N_{\alpha-1}}\frac{\sin(\sigma^{(\alpha)}_i\pm\sigma^{(\alpha-1)}_j-m^{(\alpha-1,\alpha)}_\text{bfd})}{\sin(-\sigma^{(\alpha)}_i\pm\sigma^{(\alpha-1)}_j-m^{(\alpha-1,\alpha)}_\text{bfd})}\nn\\
&&\times\prod_{k=1}^{N_{\alpha+1}}\frac{\sin(\sigma^{(\alpha)}_i\pm\sigma^{(\alpha+1)}_k-m^{(\alpha,\alpha+1)}_\text{bfd})}{\sin(-\sigma^{(\alpha)}_i\pm\sigma^{(\alpha+1)}_k-m^{(\alpha,\alpha+1)}_\text{bfd})}\times\frac{\sin^{\delta_{\alpha-1}}(\sigma^{(\alpha)}_i-m^{(\alpha-1,\alpha)}_\text{bfd})}{\sin^{\delta_{\alpha-1}}(-\sigma^{(\alpha)}_i-m^{(\alpha-1,\alpha)}_\text{bfd})}\frac{\sin^{\delta_{\alpha+1}}(\sigma^{(\alpha)}_i-m^{(\alpha,\alpha+1)}_\text{bfd})}{\sin^{\delta_{\alpha+1}}(-\sigma^{(\alpha)}_i-m^{(\alpha,\alpha+1)}_\text{bfd})}=1,\nn\\
\ea
where we set the gauge group at the $(\alpha\pm 1)$-th gauge node to be SO($2N_{\alpha\pm 1}+\delta_{\alpha\pm 1}$). For a gauge node with SO($2N_\beta+\delta_\beta$) gauge group at the $\beta$-th site for $\beta = 2,4,6,\ldots,2\lfloor\frac{r}{2}\rfloor$, we have 
\ba
\prod_{j\neq i}\frac{\sin(\sigma^{(\beta)}_i\pm \sigma^{(\beta)}_j-\beta_2\tilde{c}_\beta)}{\sin(-\sigma^{(\beta)}_i\pm \sigma^{(\beta)}_j-\beta_2\tilde{c}_\beta)}\prod_{j=1}^{N_{\beta-1}}\frac{\sin(\sigma^{(\beta)}_i\pm\sigma^{(\beta-1)}_j-m^{(\beta-1,\beta)}_\text{bfd})}{\sin(-\sigma^{(\beta)}_i\pm\sigma^{(\beta-1)}_j-m^{(\beta-1,\beta)}_\text{bfd})}\prod_{k=1}^{N_{\beta+1}}\frac{\sin(\sigma^{(\beta)}_i\pm\sigma^{(\beta+1)}_k-m^{(\beta,\beta+1)}_\text{bfd})}{\sin(-\sigma^{(\beta)}_i\pm\sigma^{(\beta+1)}_k-m^{(\beta,\beta+1)}_\text{bfd})}\nn\\
\times\frac{\sin^{\delta_\beta}(\sigma^{(\beta)}_i-\beta_2\tilde{c}_\beta)}{\sin^{\delta_\beta}(\sigma^{(\beta)}_i+\beta_2\tilde{c}_\beta)}=1.\nn\\
\ea
We leave the comparison with the Bethe ansatz equations to a future work. 

\section{Conclusion and Discussion}\label{s:dis}

In this article, we generalized the Bethe/Gauge correspondence first proposed in \cite{NS1,NS2} for A-type gauge theories to BCD-type gauge groups. We saw that the corresponding spin chain on the Bethe side is modified to one with open boundaries. In the correspondence with 2d gauge theories, we found that we can always choose diagonal boundary conditions for open XXX spin chain with the parameters $\xi$ being specified to either $\xi=0$, $\pm\frac{1}{2}$ or $\infty$ to realize the vacuum equation of gauge theory from the Bethe ansatz equation. Furthermore in the case of SO-type gauge groups, one can uplift the correspondence to a map between 3d gauge theories and open XXZ spin chain. On the other hand, when the gauge group is of Sp-type, then the straightforward uplift does not work for some reason. A similar story happens when we consider an $A_2$ quiver gauge theory with one node being SO-type and another being Sp-type, that is such a Bethe/Gauge correspondence (with diagonal open spin chain associated to the $\mathfrak{sl}_3$ R-matrix) can be established in 2d but not in 3d.

We saw that the correspondence worked perfectly for 2d gauge theories, but not as well in 3d. This might be explained in the relation to a string-theory background of these gauge theories. The brane construction of 2d gauge theories with SO and Sp type gauge groups has been given in \cite{Bergman:2018vqe} as an extension of the work \cite{Hanany:1997vm} on the construction of 2d U($N$) gauge theories. In the case of U($N$), one can use the T-duality to uplift the brane web of a 2d theory to that of a 3d theory, or even to a 4d theory. However, since the construction of SO or Sp type gauge theory involves the use of an orientifold, and the orientifold action is not preserved under the T-duality (see for example \cite{johnson_2002,Dabholkar:1997zd}), the uplift is no longer so straightforward in this case. More precisely, the orientifold action is defined as a combination of the worldsheet and the spacetime parity, while the T-duality transforms it to an operation usually denoted as $\Omega$ that reverses the left- and right-moving sectors in perturbative string theory. Interestingly, the uplift works for SO($N$) gauge theories, and it might be related to the ``trivial'' action of $\Omega$ without introducing any additional factor when exchanging the left and right Chan-Paton factors. 

For a rather similar reason, 
the 4d/2d (5d/3d) correspondence~\cite{Dorey:1998yh,Dorey:1999zk} also becomes vague after adding an orientifold. O4 (or O5) plane used in the brane construction of 4d (resp. 5d) gauge theories lies in the transverse directions to D2 (resp. D3) branes that give rise to the vortices. The effective 2d (or 3d) gauge theories on the vortices in this case is something unfamiliar to us, and it is clearly not the gauge theories with SO or Sp gauge groups considered in this article. One can also see this point by looking at the qq-characters of SO and Sp gauge theories derived in \cite{Haouzi:2020yxy} which contain infinite number of terms. The NS limit does not simplify much and the saddle-point equation of the instanton partition function in this limit appears in a different form from the Bethe ansatz equation of (diagonal) open spin chains. The quantum integrability of 4d and 5d SO and Sp gauge theories still seems to require more effort to study with better idea in the future. 

Last but not least, we recall that in the case of A-type gauge theories, starting from the TQ-relation of the corresponding periodic XXZ spin chain, 
\ba
T^A(u)Q^A(u)=\delta_+(u)Q^A(u-\eta)+e^{i\theta}\delta_-(u)Q^A(u+\eta),
\ea 
in particular when we focus on the pure gauge theory, one can rewrite it into 
\ba
\lt(\hat{y}+e^{i\theta}\hat{y}^{-1}-T^A(u)\rt)Q^A(u)=0,\quad \hat{y}:=e^{\eta\partial_u},
\ea
which matches with the spectral curve of the A-type affine Toda chain in the classical limit, where $\hat{y}$ reduces to a normal function. 
On the other hand, the expression of the eigenvalue $T(u)$ of the transfer matrix in the open XXZ spin chain can be rewritten into a TQ-relation, 
\ba
[2u]T(u)Q(u)=[2u+\eta][u+\xi_--\eta/2][u+\xi_+-\eta/2]\delta_+(u)\delta_-(-u)Q(u-\eta)\nn\\
+[2u-\eta][u-\xi_-+\eta/2][u-\xi_++\eta/2]\delta_+(-u)\delta_-(u)Q(u+\eta),
\ea
where we defined the $Q$-function as 
\ba
Q(u):=\prod_{i=1}^m[u\pm u_i].
\ea
Note that the symmetry between $u\leftrightarrow -u$ in the above TQ-relation restricts $T(u)$ to be an even function of $u$. Since the prefactors $[2u\pm\eta]$ are hard to be absorbed into the $Q$-function, it is not straightforward at all to relate the TQ-relation of the open chain to the spectral curve of the affine Toda chain of BCD-type (as it is expected in the 2d (or XXX) limit to take the form $(P(u)(\hat{y}+\mu \hat{y}^{-1})-\tilde{T}(u))Q(u)=0$ for some factor $\mu$ independent of $u$, some polynomial $P(u)$ and $\tilde{T}(u)$ some function related to $T(u)$). This might again be related to the fact that the classical limit of 4d $\cN=2$ gauge theories gives rise to the spectral curve of Toda chains \cite{Martinec:1995by}, and the relation between 4d $\cN=2$ theories and 2d theories considered in this articles is still not clear at the current stage.

\appendix 

\section*{Acknowledgment}
We would like to especially thank Hong (Kilar) Zhang for many helpful discussions at various kind of stages of this work. We also thank J. Kim,  M. Leitner, Y. Matsuo, D. O'Connor for inspiring comments and discussions on this work.
The work of TK has been supported in part by ``Investissements d'Avenir'' program, Project ISITE-BFC (No.~ANR-15-IDEX-0003), and EIPHI Graduate School (No.~ANR-17-EURE-0002). 

\section{Derivation of Bethe Ansatz Equation for XXZ Spin Chain}\label{a:bethe-open}

In this Appendix, we aim to derive Bethe ansatz equation for the open XXZ chain with the diagonal boundary operator in the form 
\ba
K(u,\xi)=\lt(\begin{array}{cc}
[u+\xi] & \\
 & -[u-\xi]\\
 \end{array}\rt).
\ea
We will first briefly review the derivation in closed chains with twisted periodic boundary condition, and then mimic it in the open chain. 

Let 
\ba
{\bf T}_0(u)={\bf R}_{0L}(u-\vartheta_L)\dots {\bf R}_{01}(u-\vartheta_1)=:\lt(\begin{array}{cc}
A(u) & B(u)\\
C(u) & D(u)\\
\end{array}\rt),
\ea
where $A(u)$, $B(u)$, $C(u)$, $D(u)\in {\rm End}(V^{\otimes L})$. By inserting the explicit expression  
\ba
{\bf T}_1(u_1){\bf T}_2(u_2)=\lt(\begin{array}{cccc}
A(u_1)A(u_2) & A(u_1)B(u_2) & B(u_1)A(u_2) & B(u_1)B(u_2)\\
A(u_1)C(u_2) & A(u_1)D(u_2) & B(u_1)C(u_2) & B(u_1)D(u_2)\\
C(u_1)A(u_2) & C(u_1)B(u_2) & D(u_1)A(u_2) & D(u_1)B(u_2)\\
C(u_1)C(u_2) & C(u_1)D(u_2) & D(u_1)C(u_2) & D(u_1)D(u_2)\\
\end{array}\rt)
\ea
and 
\ba
{\bf R}(u)=\lt(\begin{array}{cccc}
[u+\eta] &  &  & \\
 & [u] & [\eta] & \\
 & [\eta] & [u] & \\
 & & & [u+\eta]\\
\end{array}\rt),
\ea
into the RTT relation, 
\ba
{\bf R}_{12}(u_1-u_2){\bf T}_1(u_1){\bf T}_2(u_2)={\bf T}_2(u_2){\bf T}_1(u_1){\bf R}_{12}(u_1-u_2),
\ea
we obtain the commutation relations,
\begin{subequations}
\ba
&&[u_1-u_2+\eta]B(u_1)A(u_2)=[u_1-u_2]A(u_2)B(u_1)+[\eta]B(u_2)A(u_1),\\
&&[u_1-u_2+\eta]D(u_2)B(u_1)=[u_1-u_2]B(u_1)D(u_2)+[\eta]D(u_1)B(u_2),\\
&&[u_1-u_2+\eta]B(u_2)D(u_1)=[u_1-u_2]D(u_1)B(u_2)+[\eta]B(u_1)D(u_2),\\
&&\lt[u_1-u_2\rt]B(u_1)C(u_2)+\lt[\eta\rt]D(u_1)A(u_2)=\lt[u_1-u_2\rt]C(u_2)B(u_1)+\lt[\eta\rt]D(u_2)A(u_1),
\ea
\end{subequations}
and therefore, 
\ba
[u_1-u_2]D(u_2)B(u_1)=[u_1-u_2-\eta]B(u_1)D(u_2)+[\eta]B(u_2)D(u_1).
\ea

Let us consider the transfer matrix (\ref{transfer-twist}) with twisted periodic boundary condition imposed. The ground state of the system $\ket{\Omega}$ is an all-spin-up state satisfying
\ba
A(u)\ket{\Omega}=\delta_+(u)\ket{\Omega},\quad D(u)\ket{\Omega}=\delta_-(u)\ket{\Omega},\quad C(u)\ket{\Omega}=0,\label{ground-AD}
\ea
and the Bethe ansatz state is generated from $\ket{\Omega}$ by acting $B(u_i)$'s, 
\ba
\prod_{i=1}^mB(u_i)\ket{\Omega}.
\ea
The transfer matrix acts on the Bethe ansatz state as  
\ba
\lt(A(u)+e^{i\theta}D(u)\rt)B(u_1)B(u_2)\dots B(u_m)\ket{\Omega}=\prod_{i=1}^m\frac{[u_i-u+\eta]}{[u_i-u]}B(u_1)B(u_2)\dots B(u_m)A(u)\ket{\Omega}\nn\\
+e^{i\theta}\prod_{i=1}^m\frac{[u_i-u-\eta]}{[u_i-u]}B(u_1)B(u_2)\dots B(u_m)D(u)\ket{\Omega}\nn\\
-\sum_{i=1}^m \frac{[\eta]}{[u_i-u]}\prod_{j\neq i}\frac{[u_j-u_i+\eta]}{[u_j-u_i]}B(u)\prod_{j\neq i}B(u_j)A(u_i)\ket{\Omega}\nn\\
+e^{i\theta}\sum_{i=1}^m \frac{[\eta]}{[u_i-u]}\prod_{j\neq i}\frac{[u_j-u_i-\eta]}{[u_j-u_i]}B(u)\prod_{j\neq i}B(u_j)D(u_i)\ket{\Omega}+\dots
\ea
An important property here is that $\lt[B(u_i),B(u_j)\rt]=0$, which also directly follows from the RTT relation. The non-diagonal terms can be canceled against each other by imposing 
\ba
\prod_{j\neq i}[u_j-u_i+\eta]\delta_+(u_i)=e^{i\theta}\prod_{j\neq i}[u_j-u_i-\eta]\delta_-(u_i).
\ea
When we take the spin-$s_i$ representation at the $i$-th site, we have 
\ba
\delta^+(u)=\prod_{a=1}^L[u+\eta/2+\eta s_a-\vartheta_a],\quad \delta^-(u)=\prod_{a=1}^L[u+\eta/2-\eta s_a-\vartheta_a],
\ea
and thus the Bethe ansatz equation is given by 
\ba
\prod_{a=1}^L\frac{[u_i+\eta/2+\eta s_a-\vartheta_a]}{[u_i+\eta/2-\eta s_a-\vartheta_a]}=e^{i\theta}\prod_{j\neq i}^m\frac{[u_i-u_j+\eta]}{[u_i-u_j-\eta]}.\label{closed-bethe-eq}
\ea

In the case of open spin chain, we define 
\ba
U_-(u):={\bf T}(u)K(u-\frac{1}{2}\eta,\xi_-)\sigma_2{\bf T}^t(-u)\sigma_2,
\ea
which can be explicitly evaluated to 
\ba
U_-(u)=\lt(\begin{array}{cc}
\cA(u) & \cB(u)\\
\cC(u) & \cD(u)\\
\end{array}\rt),
\ea
where 
\begin{subequations}
\ba
\cA(u)=\lt[u-\frac{1}{2}\eta+\xi_-\rt]A(u)D(-u)+\lt[u-\frac{1}{2}\eta-\xi_-\rt]B(u)C(-u),\label{exp-cA}\\
\cB(u)=-\lt[u-\frac{1}{2}\eta+\xi_-\rt]A(u)B(-u)-\lt[u-\frac{1}{2}\eta-\xi_-\rt]B(u)A(-u),\\
\cC(u)=\lt[u-\frac{1}{2}\eta+\xi_-\rt]C(u)D(-u)+\lt[u-\frac{1}{2}\eta-\xi_-\rt]D(u)C(-u),\\
\cD(u)=-\lt[u-\frac{1}{2}\eta+\xi_-\rt]C(u)B(-u)-\lt[u-\frac{1}{2}\eta-\xi_-\rt]D(u)A(-u).\label{exp-cD}
\ea
\end{subequations}
We note that the ground state $\ket{\Omega}$ is also the ground state for the open chain with the diagonal boundary condition, as we have 
\ba
\cC(u)\ket{\Omega}=0.
\ea

In the same way, for open chains, we have an RURU relation given by 
\ba
{\bf R}_{12}(u_1-u_2)U_{1,-}(u_1){\bf R}_{12}(u_1+u_2-\eta)U_{2,-}(u_2)=U_{2,-}(u_2){\bf R}_{12}(u_1+u_2-\eta)U_{1,-}(u_1){\bf R}_{12}(u_1-u_2).\nn\\
\ea
What we obtain are 
\begin{subequations}
\ba
[u_1-u_2+\eta][u_1+u_2-\eta]\cB(u_1)\cA(u_2)=[\eta][u_1+u_2-\eta]\cB(u_2)\cA(u_1)\nn\\
+[u_1-u_2][u_1+u_2]\cA(u_2)\cB(u_1)
+[\eta][u_1-u_2]\cB(u_2)\cD(u_1),\label{AB-relation}\\
\lt[u_1-u_2+\eta\rt]\lt[u_1+u_2-\eta\rt]\cA(u_2)\cC(u_1)=\lt[u_1-u_2\rt]\lt[u_1+u_2\rt]\cC(u_1)\cA(u_2)\nn\\
+\lt[\eta\rt]\lt[u_1+u_2-\eta\rt]\cA(u_1)\cC(u_2)
+\lt[u_1-u_2\rt]\lt[\eta\rt]\cD(u_1)\cC(u_2),\\
\lt[u_1-u_2+\eta\rt]\lt[u_1+u_2-\eta\rt]\cD(u_2)\cB(u_1)=\lt[u_1-u_2\rt]\lt[u_1+u_2\rt]\cB(u_1)\cD(u_2)\nn\\
+\lt[\eta\rt]\lt[u_1+u_2-\eta\rt]\cD(u_1)\cB(u_2)
+\lt[u_1-u_2\rt]\lt[\eta\rt]\cA(u_1)\cB(u_2),\label{DB-relation}\\
\lt[u_1-u_2+\eta\rt]\lt[u_1+u_2-\eta\rt]\cC(u_1)\cD(u_2)=\lt[\eta\rt]\lt[u_1+u_2-\eta\rt]\cC(u_2)\cD(u_1)\nn\\
+\lt[u_1-u_2\rt]\lt[u_1+u_2\rt]\cD(u_2)\cC(u_1)
+\lt[\eta\rt]\lt[u_1-u_2\rt]\cC(u_2)\cA(u_1),
\ea
\end{subequations}
and
\begin{subequations}
\ba
&&\lt[\eta\rt]\lt[u_1+u_2\rt]\cB(u_1)\cD(u_2)+\lt[\eta\rt]^2\cA(u_1)\cB(u_2)+\lt[u_1-u_2\rt]\lt[u_1+u_2-\eta\rt]\cD(u_1)\cB(u_2)\nn\\
&&=\lt[\eta\rt]\lt[u_1-u_2+\eta\rt]\cA(u_2)\cB(u_1)+\lt[u_1+u_2\rt]\lt[u_1-u_2+\eta\rt]\cB(u_2)\cD(u_1),\label{conv-1}\\
&&\lt[\eta\rt]\lt[u_1-u_2+\eta\rt]\cB(u_1)\cD(u_2)+\lt[u_1+u_2\rt]\lt[u_1-u_2+\eta\rt]\cA(u_1)\cB(u_2)\nn\\
&&=\lt[u_1-u_2\rt]\lt[u_1+u_2-\eta\rt]\cB(u_2)\cA(u_1)+\lt[\eta\rt]^2\cB(u_2)\cD(u_1)+\lt[\eta\rt]\lt[u_1+u_2\rt]\cA(u_2)\cB(u_1).\label{conv-2}
\ea
\end{subequations}
We can further combine (\ref{AB-relation}) and (\ref{conv-2}) together to obtain 
\ba
\lt[u_1+u_2\rt]\cA(u_1)\cB(u_2)=\frac{\lt[\eta\rt]}{\lt[u_1-u_2\rt]}\lt[u_1+u_2-\eta\rt]\cB(u_1)\cA(u_2)\nn\\
+\frac{\lt[u_1-u_2-\eta\rt]}{\lt[u_1-u_2\rt]}\lt[u_1+u_2-\eta\rt]\cB(u_2)\cA(u_1)-\lt[\eta\rt]\cB(u_1)\cD(u_2),\label{conv-3}
\ea
and substitute (\ref{conv-1}) in (\ref{DB-relation}) to have   
\ba
\lt[u_1-u_2+\eta\rt]\lt[u_1+u_2-\eta\rt]\cD(u_2)\cB(u_1)=\lt[u_1-u_2\rt]\lt[u_1+u_2\rt]\cB(u_1)\cD(u_2)+\lt[u_1-u_2\rt]\lt[\eta\rt]\cA(u_1)\cB(u_2)\nn\\
+\frac{\lt[\eta\rt]\lt[u_1+u_2\rt]\lt[u_1-u_2+\eta\rt]}{\lt[u_1-u_2\rt]}\cB(u_2)\cD(u_1)+\frac{\lt[\eta\rt]^2\lt[u_1-u_2+\eta\rt]}{\lt[u_1-u_2\rt]}\cA(u_2)\cB(u_1)\nn\\
-\frac{\lt[\eta\rt]^3}{\lt[u_1-u_2\rt]}\cA(u_1)\cB(u_2)-\frac{\lt[\eta\rt]^2\lt[u_1+u_2\rt]}{\lt[u_1-u_2\rt]}\cB(u_1)\cD(u_2)\nn\\
=\lt[u_1+u_2\rt]\frac{\lt[u_1-u_2+\eta\rt]\lt[u_1-u_2-\eta\rt]}{\lt[u_1-u_2\rt]}\cB(u_1)\cD(u_2)+\lt[\eta\rt]\frac{\lt[u_1-u_2+\eta\rt]\lt[u_1-u_2-\eta\rt]}{\lt[u_1-u_2\rt]}\cA(u_1)\cB(u_2)\nn\\
+\frac{\lt[\eta\rt]\lt[u_1+u_2\rt]\lt[u_1-u_2+\eta\rt]}{\lt[u_1-u_2\rt]}\cB(u_2)\cD(u_1)+\frac{\lt[\eta\rt]^2\lt[u_1-u_2+\eta\rt]}{\lt[u_1-u_2\rt]}\cA(u_2)\cB(u_1),\nn\\
\ea
where we used 
\ba
\lt[u_1-u_2\rt]^2-\lt[\eta\rt]^2=\lt[u_1-u_2+\eta\rt]\lt[u_1-u_2-\eta\rt]. 
\ea
By further using (\ref{conv-3}), we obtain 
\ba
&&\cD(u_2)\cB(u_1)=\frac{\lt[u_1+u_2+\eta\rt]\lt[u_1-u_2-\eta\rt]}{\lt[u_1-u_2\rt]\lt[u_1+u_2\rt]}\cB(u_1)\cD(u_2)+\frac{\lt[\eta\rt]\lt[u_1+u_2+\eta\rt]}{\lt[u_1-u_2\rt]\lt[u_1+u_2\rt]}\cB(u_2)\cD(u_1)\nn\\
&&+2\lt[\eta\rt]^2\frac{\cosh(\eta)}{\lt[u_1-u_2\rt]\lt[u_1+u_2\rt]}\cB(u_1)\cA(u_2)+\lt[\eta\rt]\frac{\lt[u_1-u_2-2\eta\rt]}{\lt[u_1-u_2\rt]\lt[u_1+u_2\rt]}\cB(u_2)\cA(u_1),\label{DB-p}
\ea
where we used 
\ba
\lt[u_1-u_2-\eta\rt]+\lt[u_1-u_2+\eta\rt]=2[u_1-u_2]\cosh(\eta).
\ea
We define a convenient notation instead of $\cD(u)$, 
\ba
\tilde{\cD}(u)=[2u]\cD(u)-[\eta]\cA(u),
\ea
which following from (\ref{DB-p}) satisfies 
\ba
\tilde{\cD}(u_2)\cB(u_1)=\frac{\lt[u_1+u_2+\eta\rt]\lt[u_1-u_2-\eta\rt]}{\lt[u_1-u_2\rt]\lt[u_1+u_2\rt]}\cB(u_1)\tilde{\cD}(u_2)+\frac{\lt[\eta\rt]\lt[2u_2+\eta\rt]}{\lt[u_1-u_2\rt]\lt[2u_1\rt]}\cB(u_2)\tilde{\cD}(u_1)\nn\\
+\frac{\lt[\eta\rt]\lt[2u_1-\eta\rt]\lt[2u_2+\eta\rt]}{\lt[u_1+u_2\rt]\lt[2u_1\rt]}\cB(u_2)\cA(u_1),
\ea
and one can also rewrite (\ref{AB-relation}) to 
\ba
&&\cA(u_2)\cB(u_1)=\frac{\lt[u_1-u_2+\eta\rt]\lt[u_1+u_2-\eta\rt]}{\lt[u_1-u_2\rt]\lt[u_1+u_2\rt]}\cB(u_1)\cA(u_2)-\frac{\lt[\eta\rt]\lt[2u_1-\eta\rt]}{\lt[u_1-u_2\rt]\lt[2u_1\rt]}\cB(u_2)\cA(u_1)\nn\\
&&-\frac{\lt[\eta\rt]}{\lt[u_1+u_2\rt]\lt[2u_1\rt]}\cB(u_2)\tilde{\cD}(u_1).
\ea

By using (\ref{ground-AD}), we have
\begin{subequations}
\begin{align}
 \cA(u)\ket{\Omega}&=\lt[u-\frac{1}{2}\eta+\xi_-\rt]\delta_+(u)\delta_-(-u),\label{A-action}\\
\cD(u)\ket{\Omega}
&=\lt[u-\frac{1}{2}\eta+\xi_-\rt]\frac{[\eta]}{[2u]}\delta_+(u)\delta_-(-u)-\lt[u+\frac{1}{2}\eta-\xi_-\rt]\frac{[2u-\eta]}{[2u]}\delta_+(-u)\delta_-(u),
\end{align}
\end{subequations}
which implies 
\ba
\tilde{\cD}(u)\ket{\Omega}&=&-\lt[u+\frac{1}{2}\eta-\xi_-\rt][2u-\eta]\delta_+(-u)\delta_-(u)\label{tD-action},
\ea
and thus the transfer matrix at ground state is evaluated to 
\ba
\bra{\Omega}\mathsf{t}(u)\ket{\Omega}=\lt[u-\frac{1}{2}\eta+\xi_+\rt]\lt[u-\frac{1}{2}\eta+\xi_-\rt]\frac{[2u+\eta]}{[2u]}\delta_+(u)\delta_-(-u)\nn\\
+\lt[u+\frac{1}{2}\eta-\xi_+\rt]\lt[u+\frac{1}{2}\eta-\xi_-\rt]\frac{[2u-\eta]}{[2u]}\delta_+(-u)\delta_-(u).
\ea
Note that the above transfer matrix is symmetric about $u\leftrightarrow -u$. We rewrite $\mathsf{t}(u)$ in terms of $\tilde{D}(u)$, 
\ba
\mathsf{t}(u)=[u+\frac{\eta}{2}+\xi_+]\cA(u)-[u+\frac{\eta}{2}-\xi_+]\cD(u)=\frac{\lt[2u+\eta\rt]\lt[u+\xi_+-\frac{\eta}{2}\rt]}{\lt[2u\rt]}\cA(u)-\frac{\lt[u+\frac{\eta}{2}-\xi_+\rt]}{\lt[2u\rt]}\tilde{\cD}(u).\nn\\
\ea

For excited Bethe states, we need to consider 
\ba
&&\lt(\lt[2u+\eta\rt]\lt[u+\xi_+-\frac{\eta}{2}\rt]\cA(u)-\lt[u+\frac{\eta}{2}-\xi_+\rt]\tilde{\cD}(u)\rt)\cB(u_1)\cB(u_2)\dots \cB(u_k)\ket{\Omega}\nn\\
&&=\lt[2u+\eta\rt]\lt[u+\xi_+-\frac{\eta}{2}\rt]\prod_{i=1}^k\frac{[u_i-u+\eta][u+u_i-\eta]}{[u_i-u][u+u_i]}\cB(u_1)\cB(u_2)\dots \cB(u_k)\cA(u)\ket{\Omega}\nn\\
&&-[u+\frac{\eta}{2}-\xi_+]\prod_{i=1}^k\frac{[u_i-u-\eta][u_i+u+\eta]}{[u_i-u][u_i+u]}\cB(u_1)\cB(u_2)\dots \cB(u_k)\tilde{\cD}(u)\ket{\Omega}\nn\\
&&-\lt[2u+\eta\rt]\lt[u+\xi_+-\frac{\eta}{2}\rt]\sum_{i=1}^k\frac{[\eta][2u_i-\eta]}{[u_i-u][2u_i]}\prod_{j\neq i}\frac{[u_j-u_i+\eta][u_i+u_j-\eta]}{[u_j-u_i][u_i+u_j]}\cB(u)\prod_{j\neq i}\cB(u_j)\cA(u_i)\ket{\Omega}\nn\\
&&-\lt[2u+\eta\rt]\lt[u+\xi_+-\frac{\eta}{2}\rt]\sum_{i=1}^k\frac{[\eta]}{[2u_i][u_i+u]}\prod_{j\neq i}\frac{[u_j-u_i-\eta][u_j+u_i+\eta]}{[u_j-u_i][u_j+u_i]}\cB(u)\prod_{j\neq i}\cB(u_j)\tilde{\cD}(u_i)\ket{\Omega}\nn\\
&&-\lt[u+\frac{\eta}{2}-\xi_+\rt]\sum_{i=1}^k\frac{[\eta]\lt[2u+\eta\rt]}{\lt[u_i-u\rt]\lt[2u_i\rt]}\prod_{j\neq i}\frac{[u_j-u_i-\eta][u_j+u_i+\eta]}{[u_j-u_i][u_j+u_i]}\cB(u)\prod_{j\neq i}\cB(u_j)\tilde{\cD}(u_i)\ket{\Omega}\nn\\
&&-\lt[u+\frac{\eta}{2}-\xi_+\rt]\sum_{i=1}^k\frac{[\eta]\lt[2u_i-\eta\rt]\lt[2u+\eta\rt]}{[2u_i][u_i+u]}\prod_{j\neq i}\frac{[u_j-u_i+\eta][u_i+u_j-\eta]}{[u_j-u_i][u_i+u_j]}\cB(u)\prod_{j\neq i}\cB(u_j)\cA(u_i)\ket{\Omega}+\dots\nn\\
\ea
Therefore the Bethe ansatz equation is given by 
\ba
&&\lt(\frac{\lt[u+\frac{\eta}{2}-\xi_+\rt]\lt[2u+\eta\rt]\lt[\eta\rt]}{\lt[u_i-u\rt]\lt[2u_i\rt]}+\frac{\lt[2u+\eta\rt]\lt[u+\xi_+-\frac{\eta}{2}\rt]\lt[\eta\rt]}{\lt[2u_i\rt]\lt[u+u_i\rt]}\rt)\prod_{j\neq i}\frac{[u_j-u_i-\eta][u_j+u_i+\eta]}{[u_j-u_i][u_j+u_i]}\tilde{\cD}(u_i)\ket{\Omega}\nn\\
&&+\lt(\frac{\lt[2u+\eta\rt]\lt[u+\xi_+-\frac{\eta}{2}\rt]\lt[2u_i-\eta\rt]\lt[\eta\rt]}{\lt[u_i-u\rt]\lt[2u_i\rt]}+\frac{\lt[u+\frac{\eta}{2}-\xi_+\rt][\eta]\lt[2u_i-\eta\rt]\lt[2u+\eta\rt]}{[2u_i][u_i+u]}\rt)\nn\\
&&\times\prod_{j\neq i}\frac{[u_j-u_i+\eta][u_i+u_j-\eta]}{[u_j-u_i][u_i+u_j]}\cA(u_i)\ket{\Omega}=0,
\ea
and it can be simplified to 
\ba
\frac{\lt[2u_i-\eta\rt]\lt[u_i+\xi_+-\frac{\eta}{2}\rt]}{\lt[u_i-\xi_++\frac{\eta}{2}\rt]}\prod_{j\neq i}\frac{[u_j-u_i+\eta][u_i+u_j-\eta]}{[u_j-u_i-\eta][u_j+u_i+\eta]}\cA(u_i)\ket{\Omega}=-\tilde{\cD}(u_i)\ket{\Omega}.
\ea
By substituting the explicit expression of $\cA(u)\ket{\Omega}$ and $\tilde{\cD}(u)\ket{\Omega}$, (\ref{A-action}) and (\ref{tD-action}) into the above equation, we obtain the final result of the Bethe ansatz equation, 
\ba
\frac{\lt[u_i+\xi_+-\frac{\eta}{2}\rt]\lt[u_i-\frac{\eta}{2}+\xi_-\rt]\delta_+(u_i)\delta_-(-u_i)}{\lt[u_i-\xi_++\frac{\eta}{2}\rt]\lt[u_i+\frac{\eta}{2}-\xi_-\rt]\delta_+(-u_i)\delta_-(u_i)}\prod_{j\neq i}\frac{[u_j-u_i+\eta][u_i+u_j-\eta]}{[u_j-u_i-\eta][u_j+u_i+\eta]}=1.
\ea

\bibliography{MO-Bethe}

\end{document}